# Improving the State of the Art for Training Human-AI Teams

**Technical Report #1**

# Results of Subject-Matter Expert Knowledge Elicitation Survey

**March 2023**

*Jim McCarthy, Lillian Asiala, LeeAnn Maryeski, Nyla Warren*

*Sonalysts, Inc.*

This Page Intentionally Blank

# TABLE OF CONTENTS

# LIST OF FIGURES

# LIST OF TABLES

This Page Intentionally Blank

# ACRONYMS

**A**
AFRL ..................................................................................................*Air Force Research Laboratory*
AI .................................................................................................................*Artificial Intelligence*

**G**
GUI ..........................................................................................................*Graphical User Interface*

**I**
IAMD...............................................................................................*Integrated Air and Missile Defense*

**J**
JADC2..........................................................................................*Joint All-Domain Command and Control*

**N**
NASEM ............................................................*National Academies of Sciences, Engineering, and Medicine*
NIMS ........................................................................................*National Incident Management System*

**O**
OODA..................................................................................................*Observe-Orient-Decide-Act*

**S**
SA ..............................................................................................................*Situation Awareness*
SATCOMS.................................................................................................*Satellite Communications*
SME .........................................................................................................*Subject-Matter Expert*
STE..........................................................................................................*Synthetic Task Environment*

**T**
TIDE ................................................................................................*Targeted Ideation Development Event*

**V**
VoIP.........................................................................................................*Voice over Internet Protocol*

This Page Intentionally Blank

# ACKNOWLEDGEMENTS

We would like to acknowledge the subject-matter experts who donated their time and expertise to help us develop and complete this survey:

                                        Bob Banker

                                        John Dickmann

                                        Al Elkins

                                        Bill Glenney

                                        Dixon Hicks

                                        Kurush Morris

                                        Nicholas Raic

We would also like to express our appreciation to those participants who asked to remain anonymous.

This Page Intentionally Blank

# Abstract

A consensus report produced for the Air Force Research Laboratory by the National Academies of Sciences, Engineering, and Mathematics documented a prevalent and increasing desire to support human-Artificial Intelligence (AI) teaming across military service branches. Sonalysts has begun an internal initiative to explore the training of human-AI teams. The first step in this effort is to develop a Synthetic Task Environment (STE) that is capable of facilitating research on human-AI teams. We decided to use Joint All-Domain Command and Control (JADC2) as a focus point for developing the STE because the volume of sensor inputs and decision options within the JADC2 concept likely requires the use of AI systems to enable timely decisions. Given this focus, we engaged a number of Subject-Matter Experts (SMEs) with Command and Control experience to gain insight into developing a STE that embodied the teaming challenges associated with JADC2. This report documents our initial engagement with those stakeholders. We created a survey that featured two types of questions. The first asked respondents to report their agreement with the STE features that we anticipated might be important within a JADC2-focused testbed. The second type asked SMEs to respond to open-ended questions exploring testbed features such as the task domain, methods of performance assessment, methods of communication, and characteristics of autonomous teammates. The research team identified thirteen Sonalysts employees (referred to internally as partners) with military backgrounds and Command and Control experience, and invited them to participate. Twelve respondents completed the survey. The team then analyzed the responses to identify themes that emerged and topics that would benefit from further analysis. The results indicated that our SMEs were amenable to research using tasks that were analogous to those encountered in military environments, as long as they required teams to process a great deal of incoming data to arrive at complex decisions. The SMEs felt that the testbed should support "teams of teams" that represented a matrixed organization, and that it should support a robust array to spoken, text-based, and face-to-face communications.

This Page Intentionally Blank

# 1 BACKGROUND

In 2021, the United States Air Force Research Laboratory (AFRL) Human Performance Wing asked the National Academies of Sciences, Engineering, and Medicine (NASEM) to produce a consensus report to examine the military role of Artificial Intelligence (AI), particularly as part of Human-AI teams. The goal of this effort was to allow AFRL to better support the design of future systems in which humans are teamed with intelligent agents to achieve mission objectives.

The NASEM report identified nine focus areas within the broader Human-AI-teaming domain:

1. Training Human-AI Teams
2. AI Transparency and Explainability
3. Human-AI Team Interaction
4. Trusting AI Teammates
5. Human-AI Teaming Processes and Effectiveness
6. Human-AI Teaming Methods and Models
7. Situation Awareness in Human-AI Teams
8. Identification and Mitigation of Bias in Human-AI Teams
9. Human-System Integration Processes and Measures of Human-AI Team Collaboration and Performance

Sonalysts has begun an internal initiative to explore the first of these Human-AI-teaming domains – Training Human-AI Teams. Table 1 reproduces a portion of the NASEM report in which the authors distributed the six research needs over three periods.

Table 1: NASEM Training Research Needs Over Time

|  | Near-Term (1-5 Years) | Mid-Term (6-10 Years) | Far-Term (11-15 Years) |
|---|---|---|---|
| **Training Human-AI Teams** | 9.1: Developing Human-Centered Human-AI Team-Training Content | | |
| | 9.2: Testing and Validating Traditional Team Training Methods to Inform New Methods | | |
| | | 9.3: Training to Calibrate Human Expectations of Autonomous Teammates | |
| | 9.4: Designing Platforms for Human-AI Team Training | | |
| | | 9.5: Adaptive Training Materials Based on Differing Team Compositions and Sizes | |
| | | 9.6: Training That Works Toward Trust in Human-AI Teaming | |

The first step in our research plan is the development of a Synthetic Task Environment (STE) that will provide a validated research environment in which our Human-AI teams can perform. To arrive at a set of requirements that will maximize the applicability of our research, Sonalysts is engaging with leading researchers and Subject-Matter Experts (SMEs) within the field. This report summarizes the findings of the first of these outreach efforts with military SMEs.

## 2 Overview

Although Human-AI teaming is of interest across multiple echelons of each military service, a focusing lens is provided by emerging Joint All-Domain Command and Control (JADC2) concepts. The JADC2 concept envisions sensors from all military services and all domains (*e.g.,* air, sea, land, space, cyberspace) being connected in a vast network to enable rapid employment of assets across those services and domains to achieve mission effects. A significant goal is to make combat decisions and take the corresponding actions much more rapidly than is currently possible (*i.e.,* reduce typical timelines from days to hours). To increase operating tempo while simultaneously considering a much greater array of data and potential Courses of Action (COA), it is likely that human decision-makers will be teamed with AI teammates much more extensively than has occurred in the past. Further, we envision that this will be true "teaming." Our research will assume that autonomous agents will eventually be able to work with humans as peers/teammates, not merely tools. To achieve this, we envision that Human-AI Teams will involve **autonomous systems** that use AI or similar technologies to make decisions and/or take actions within a specified domain. These systems will be capable of responding to novel performance challenges while, coordinating, and cooperating with teammates. When these milestones are reached, we can begin to view the system as an autonomous teammate that is capable of not only independent action, but also **interdependent** action.

## 3 Survey Development and Analysis

Given the importance of JADC2 to the future of operations and the likelihood that JADC2 will lead to the formation of Human-AI teams, we felt it was prudent to involve SMEs in the development of the STE. Sonalysts' goal was to ensure that the STE embodied the type of teamwork, tasks, and decision-making that are likely to be encountered in a JADC2 environment, while creating a problem domain that was unclassified and accessible to those with little or no military background. Our goal was **NOT** to explore specific JADC2 concepts/technologies.

Using our research plan as a foundation, the research team identified various topics for which SMEs might be able to provide useful inputs. Working from these knowledge elicitation objectives, the team then developed and reviewed a range of open-ended and Likert-style questions that would allow us to gather insights regarding the beliefs and priorities of our SME sample. Following our internal quality assurance process, the lead researchers "authored" the survey within the Survey Monkey delivery tool. Prior to releasing the survey, we conducted a pilot test during which a SME completed the survey while a member of the research team observed. The pilot test led to some minor refinements, mostly to the directions provided to the respondents. For example, because JADC2 is an emerging concept, it was likely that many of our respondents would lack direct familiarity with this environment. To address this, the research team updated the directions to soften the focus on JADC2 and direct respondents to use their experience with military planning, command, and control as a useful analog. Researchers reinforced this message in any direct discussions with the respondents.

Following the pilot test, the research team released the survey to 13 senior SMEs. Each SME received a unique link that allowed the respondent to complete the survey across several sessions, if desired. The unique links also allowed us to monitor the progress of respondents as they completed the survey. The lead researchers provided reminder emails to encourage a high response rate. In the end, 12 respondents completed the entire survey.

When the response period ended, the research team downloaded the results from Survey Monkey for offline analysis. While many respondents seemed focused on developing an environment that would allow the evaluation of JADC2 concepts (rather than an environment that we could use to evaluate Human-AI teaming, albeit with a family resemblance to JADC2), we did receive useful inputs across the survey. The inputs were useful in their own right, and set the stage for more focused discussions in the future.

The analysis of the survey results occurred in two threads. First, the two lead researchers conducted independent thematic analyses of the responses for each open-ended question. Within these analyses, each researcher reviewed each respondent's answer to each question and captured it as a short phrase/theme. The goal was to identify ideas that recurred across answers, even if respondents used different wording. After completing the independent analyses, the researchers established a consensus list of themes for each question and mapped the individual responses to that consensus list.

Second, in parallel with this qualitative analysis of the open-ended questions, the research team also conducted a quantitative analysis of the various Likert-like items. These analyses focused on descriptive statistics summarizing the distribution of responses across the 12 respondents.

The lead researchers then documented the survey findings, and our interpretation of those findings, in this technical report. The report underwent an internal quality assurance process, and was then archived and distributed to stakeholders (including survey respondents).

The following sections summarize the inputs we received as they pertain to various elements of the synthetic task environment. The raw survey responses are provided in Appendix A.

## 4 DISCUSSION OF PROBLEM DOMAIN

A key element of the STE is the task domain. The task domain defines the mission of the teams and provides the context for the tasks that they will perform. While we want the task domain to require the same types of information processing, decision-making, and coordination that is likely to be required in a JADC2 environment, using an analog environment may make it easier to create authentic tasks while remaining unclassified, opening up a broader pool of participants.

This portion of the survey focused on identifying the features of the task domain that would enhance the credibility and applicability of our research to real-world operations.

### 4.1 Analysis of Open-Ended Items

Two sets of questions had relevance to the definition of the problem domain and the tasks presented to users within the domain. Because we wanted to investigate the suitability of JADC2-like tasks for the STE, one set of questions focused on that domain. These questions were:

1. How are JADC2 teams structured?
2. Generically, what types of tasks are performed in a JADC2 environment?
3. Is the Observe-Orient-Decide-Act (OODA) loop a good metaphor for how JADC2 operations are conducted? If not, what is a better metaphor?
4. Are there hierarchical relationships in JADC2 teams (*i.e.*, a clear command hierarchy) or are team members typically more parallel in authority? Does this change based on tasking?

5. Do you have any remaining ideas, comments, or questions about JADC2 operations? If so, please record them here.

Table 2 summarizes the themes that appeared in the respondents' answers to these questions.

Table 2: Analysis of Open-ended Responses to JADC2-focused Questions

| Question | Analysis |
|---|---|
| How are JADC2 teams structured? | The most common answer (5 of 12 respondents) was some variation of "I don't know." <br><br> Another frequent response (n=5) was some variation of a matrix in which the columns were various operational domains (sea, air, land, *etc*.) and the rows were various functions (common operating picture, missile defense, emergency action controller, intelligence, *etc*.) <br><br> We encouraged respondents to use their experience in planning, command, and control functions as a rough analog for JADC2 operations, if they lacked direct experience. This guidance generated a number of responses that were each produced by a single respondent. These responses included items such as similarity to the joint staff; similarity to the U.S. Combined Arms Operations Structure; and similarity to existing Command and Control structures, as dictated by the lead service and variably (as dictated by the nature of the task and the available resources). |
| Generically, what types of tasks are performed in a JADC2 environment? | Two answers to this prompt were submitted by several respondents. The first was some variation of "sensor fusion" (n=4) and the second touched on the theme of "COA decision-making" (n=3). Two other themes were suggested by two respondents each: <br><br> 1. Maintaining/promulgating Situation Awareness (SA) and Common Operating Picture <br> 2. Multi-level coordination (including prioritization and deconfliction) <br><br> Another group of responses focused on the conventional "kill chain" (Find, Fix, Track, Target, Engage, and Assess). One respondent identified all the phases; a second respondent focused on Tracking and Targeting. <br><br> Individual respondents provided a number of responses. These included battle management of assets, domain awareness, planning, and status reporting. |

| Question | Analysis |
|---|---|
| Is the OODA loop a good metaphor for how JADC2 operations are conducted? If not, what is a better metaphor? | Seven of the twelve respondents thought that the OODA loop served as a good metaphor, with individuals specifically calling out its applicability at the theater level (n=1) and for active operations (n=1).<br><br>Two respondents explicitly said "no." These two respondents offered suggestions for better metaphors. Specifically, one respondent suggested the use of link-node networks and another suggested the use of the six-phase joint targeting process.<br><br>The remaining three respondents said that they did not know (n=1) or that it was possible (n=2). |
| Are there hierarchical relationships in JADC2 teams (*i.e.*, a clear command hierarchy) or are team members typically more parallel in authority? Does this change based on tasking? | Half of the respondents (n=6) felt that JADC2 teams were hierarchically organized. Three additional respondents expressed that teams were mostly parallel within the domain/function matrix, but with a clear director.<br><br>Three respondents indicated that they did not know enough to offer an opinion. One respondent noted that JADC2 might lead to an increase in centralized decision-making. |
| Do you have any remaining ideas, comments, or questions about JADC2 operations? If so, please record them here. | This question did not produce any useful replies. Some responses indicated that the participants were focused on defining JADC2 and/or testing JADC2 concepts, not on defining a STE that resembled JADC2 organizations and processes, albeit in a different context. |

The second set of questions explored the important features of the problem domain and the included task. Those four questions were:

1. What is a good analogous task domain for the STE to incorporate (*i.e.*, a domain that is analogous to the JADC2 task domain without being classified)?
2. What qualities would that domain need to possess?
3. Do you have thoughts about the "game" that participants should be asked to play within the STE?
4. Do you have any remaining ideas, comments, or questions about the task domain used in the STE? If so, please record them here.

Table 3 captures our thematic analysis of the responses to those items.

**Table 3: Analysis of Open-ended Responses to Task Attribute Questions**

| Question | Analysis |
|---|---|
| What is a good analogous task domain for the STE to incorporate (*i.e.*, a domain that is analogous to the JADC2 task domain without being classified)? | There was no specific overlap among the responses to this question. However, several responses adopted an approach similar to, "any task that emphasizes maintaining SA and responding as conditions warrant" (n=1). The responses that had a "family resemblance" to this notion were (n=1, in all cases):<br><br>• Crisis/Incident response, perhaps embodying the National Incident Management System (NIMS)<br>• A paintball, capture the flag type of team competition that incorporates a command element with field forces executing tasks<br>• Mountain/wilderness/open ocean search and rescue<br>• Integrated Air and Missile Defense (IAMD)<br>• Amphibious Landing<br><br>We also received three other responses that did not fit neatly within this theme (n=1, in all cases):<br><br>• Uber Model (https://sgp.fas.org/crs/natsec/IF11493.pdf)<br>• Tasker System<br>• Something based on 9/11<br><br>One respondent simply said that they could not provide a useful answer. Another suggested that we leverage the Universal Joint Task List. |

| Question | Analysis |
|---|---|
| What qualities would that domain need to possess? | The research team received good input here, but with little thematic overlap.<br><br>Two respondents noted that the task domain needed to include rapidly evolving scenarios with multiple simultaneous axes of action and decision-making (n=2) that require coordination (n=1). Similarly, respondents noted that the task should provide an overwhelming amount of data to decision-makers (n=1) and that the level of information/clarity must be controllable to mimic the fog of war (n=1). Given our focus on JADC2-like tasks, respondents encouraged us to ensure that the task required the involvement of different organizations (n=1) and/or warfare in several domains (n=1).<br><br>A number of responses discussed various aspects of the task that should populate the STE. For example, respondents identified several task types that should be included, such as (n=1, unless otherwise noted):<br><br>- Communication<br>- Asset/Resource management<br>- Personnel management<br>- Sensor management and real-time intelligence collection (n=2)<br>- Planning<br><br>Several other inputs were more focused on features of the task environment (n=1, in all cases). This included responses that indicated that the task environment should:<br><br>- Include surprise and/or hostile acts<br>- Be repeatable<br>- Be scalable<br>- Be structured<br>- Require timeliness<br>- Require responsiveness<br><br>Three other responses resisted grouping with other inputs (n=1, in all cases):<br><br>- Knowledge of direct and indirect forcing functions within the domain environment<br>- Lack of ambiguity about who makes decisions<br>- Results in milestone-based decision-making |

| Question | Analysis |
|---|---|
| Do you have thoughts about the "game" that participants should be asked to play within the STE? | In addition to input that was provided earlier (*e.g.*, build the STE around a paintball or capture the flag type of team competition that incorporates a command element with field forces executing tasks), respondents noted that it would be important to provide time-critical tasks (n=2) that required players to prioritize and multitask (n=1). The tasks should also have meaningful decision consequences (n=1) and present a spectrum of winning and losing possibilities (n=1). |
| Do you have any remaining ideas, comments, or questions about the task domain used in the STE? If so, please record them here. | The research team received three responses to this item that echoed themes established elsewhere (n=1, in all cases):<br>• Ensure that there is pressure to win<br>• Ensure a continuous data stream with meaningful and meaningless inputs to challenge SA and re-planning<br>• Data availability should be an important independent variable |

### 4.2 Assessment of Discrete Features

Preceding the survey, Sonalysts conducted a fairly broad literature review. The review, coupled with our experience in a range of modeling and simulation efforts, allowed us to predict a range of task domain features that we thought might be useful to include. Within the survey, we asked respondents to indicate the degree to which they agreed or disagreed with our hypotheses about which specific features would be important within the STE. Pie charts and descriptive statistics for each of the domain-related features are shown in Appendix B.

In this section, we focus attention on the items with the greatest difference of opinion among the respondents. These items may represent topics that would benefit from further discussion. To establish a reasonable basis of comparison across the various sections, the research team decided to use a standard deviation of greater than or equal to 1.0 as our threshold. The team flagged items at or above that threshold for further consideration. Four of the 11 items associated with this section met that standard.

One of the four items garnered a wide spread of responses: "The task environment must require the same type of data gathering, decision-making, and action-taking that characterizes any military operation that employs the OODA loop." Figure 1 shows the spread of agreement associated with this feature (standard deviation =1.20). For this item, the majority of responses were neutral or positive, with one respondent indicating strong disagreement with the item.

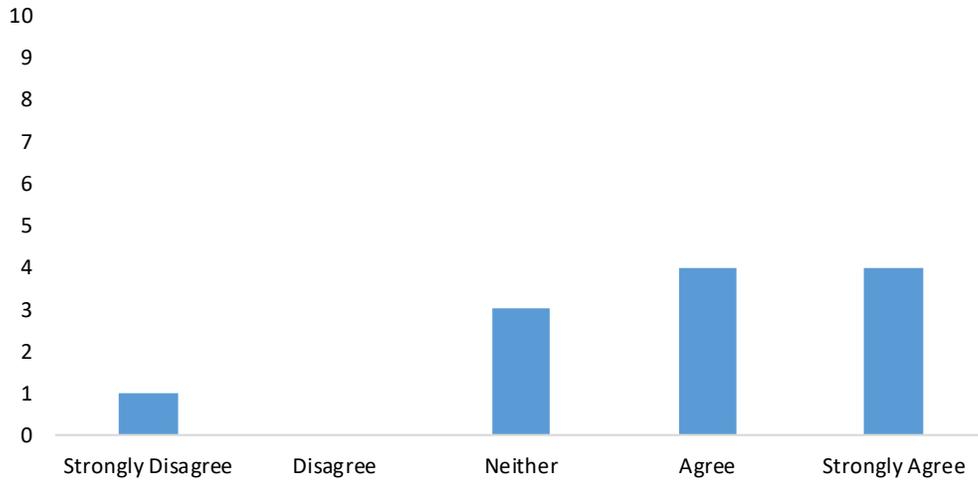

**Figure 1. Agreement ratings for "The task environment must require the same type of data gathering, decision-making, and action-taking that characterizes any military operation that employs the Observe-Orient-Decide-Act (OODA) loop."**

A second feature that generated a spread of responses was "The task environment must support the ability to modulate task uncertainty." Agreement ratings for this feature are displayed in Figure 2. This range was more evenly distributed over the "disagree" options (standard deviation = 1.38), but the most frequent response was "strongly agree."

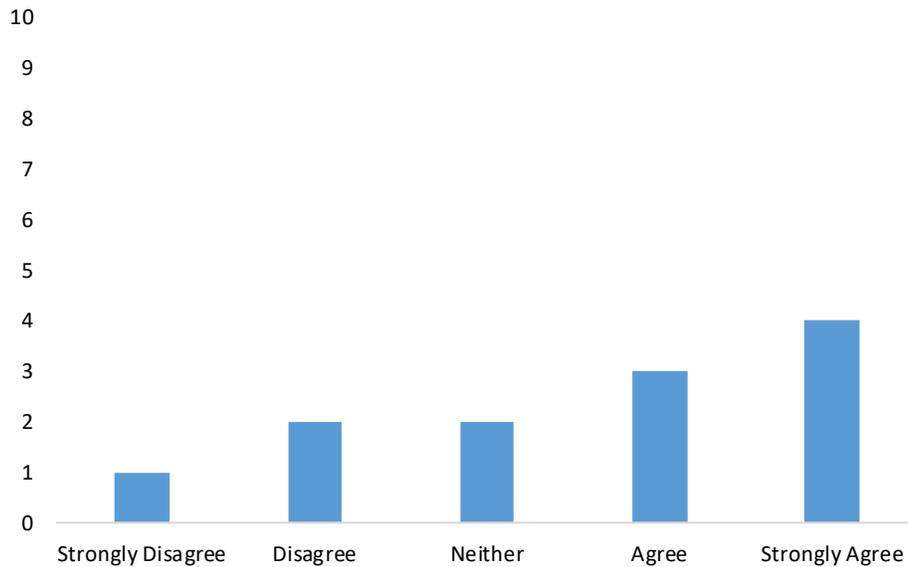

**Figure 2. Agreement ratings for "The task environment must support the ability to modulate task uncertainty."**

Another item, similar to the above, with a significant distribution of responses was "The tasks in the STE should have different levels of uncertainty." Responses to this item were distributed across "disagree" to "strongly agree," with "strongly agree" being the most frequent response (see Figure 3).

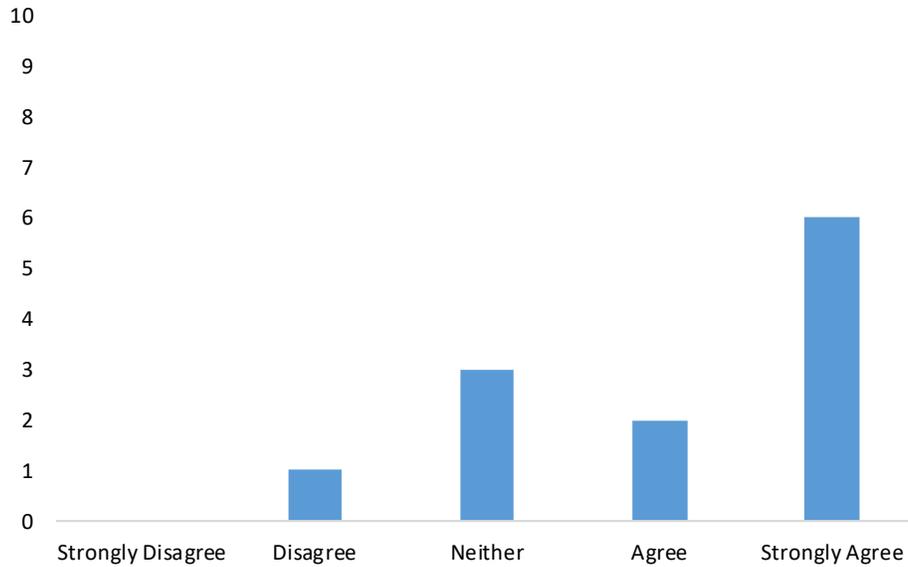

**Figure 3. Agreement Ratings for "The tasks in the STE should have different levels of uncertainty."**

The fourth feature with a wide range of responses was "The tasks in the STE should have different 'stakes' (and apply different levels of pressure on operators)." Responses to this feature can be found in Figure 4. Most respondents "strongly agreed" that this should be a feature incorporated in the STE, with the second most popular response being "neither agree nor disagree." One respondent felt strongly that this should not be incorporated as a feature. The standard deviation for this item was 1.30.

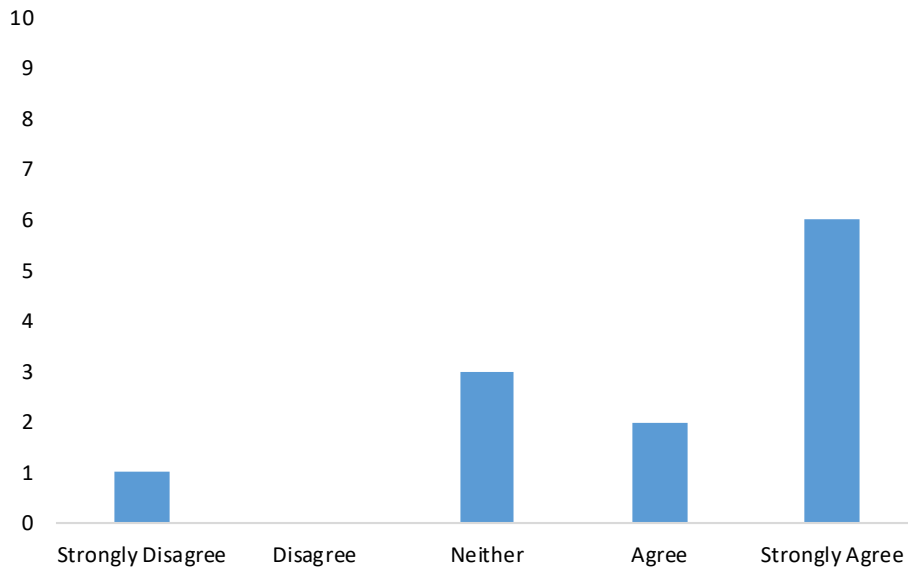

**Figure 4. Agreement ratings for "The tasks in the STE should have different 'stakes' (and apply different levels of pressure on operators)."**

### 4.3 Summary of Findings

Across a number of open-ended questions, we assessed the qualities of the task domain that are likely to be important. In reviewing the responses, the themes below began to emerge.

The domain should require teams of teams in which each sub-team focused on a specific aspect of the problem and there are multiple roles within each team. A structure like the U.S. Arms Operations or Joint Staff Structure might be useful. Each sub-team should operate in a largely parallel fashion, with their respective inputs filtering up to a central decision authority. The various roles/functions within a sub-team largely operate in parallel.

One specific suggestion for an analog problem domain was something modeled on paint ball or capture-the-flag. The respondent suggested this approach because designers could structure the game to include information gathering, command tasks, and field-level execution of plans. Another suggestion was crisis/incident response modeled on the operation of the NIMS. Others suggested a wilderness or open-ocean search-and-rescue type of scenario. Overall, these suggestions were consistent with a STE feature that had high agreement among respondents (*i.e.,* respondents either "agreed" or "strongly agreed"), specifically that the task environment must require the type of coordination and decision-making anticipated within a JADC2 context ($\bar{x} = 4.50$).

The SMEs felt that the tasks presented to teams within the task domain should focus on information gathering from a wide collection of sources. Teams must be required to sort relevant from irrelevant data to develop, maintain, and promulgate the current situation assessment/common operating picture. The tasks should include operational planning, sensor management (including tracking/targeting functions), asset/resource management, personnel management, communication, and the selection among COAs. This process should require multi-level coordination, including prioritization, deconfliction, and multi-tasking. The task environment should include rapidly evolving scenarios, time-critical tasks, and opportunities for surprise, random occurrences, and/or hostile acts. There should be a clear focus on "winning," and researchers should throttle the availability of data to control difficulty/uncertainty. More generally, the task environment should be structured, scalable, and repeatable. Finally, the task presented in the task domain should present variable levels of difficulty.

Additional features with significant agreement were:

1. The task environment must provide a venue for practicing and assessing coordination behaviors across the timeline (*i.e.,* from "preparation" activities to "adjustment" activities; $\bar{x} = 4.33$).
2. The task environment must include opportunities for mission planning, mission execution, and mission debrief ($\bar{x} = 4.33$).
3. The tasks in the STE should have different levels of time pressure ($\bar{x} = 4.33$).
4. The tasks in the STE should have different levels of difficulty ($\bar{x} = 4.25$).

## 4.4 Topics for Further Exploration

Our research plan includes opportunities to follow-up with survey respondents to understand their inputs more clearly and work toward consensus on issues of apparent disagreement. With respect to the task domain, it may be useful to engage SMEs in a more complete discussion of:

- The role of the OODA loop within JADC2 and analogous teams
- Whether or not to incorporate different difficulty factors (*i.e.,* stakes/levels of uncertainty) in the STE
- The team structure(s) that should be supported within the STE

It may also be useful to present a notional task environment and ask the SMEs to react to and improve it.

# 5 DISCUSSION OF PERFORMANCE ASSESSMENT AND DATA COLLECTION

Two related issues that will shape the development of the STE are the nature of the data that should be collected within the STE and the performance evaluation options that should be supported. A core capability of the STE must be rich data collection and performance assessment capabilities. The STE must be able to assess and store mission outcomes, task outcomes, and detailed task processes. Pending further definition, we plan to emphasize data collection and evaluation that is real-time, automatic, and nonobtrusive.

Both of these issues are of primary concern to the research community (and thus explored in a separate knowledge elicitation survey, see *Technical Report 2: Improving the State of the Art for Training Human-AI Teams – Results of Researcher Knowledge Elicitation Survey*). However, we felt that useful insights could be gained by examining how performance is captured/assessed within the JADC2 environment or similar planning, command, and control domains. In this section, we document the lessons learned within the related portions of the survey.

## 5.1 Analysis of Open-Ended Items

The survey included eight items related to data capture and performance evaluation:

1. What kind of performance variables do you think would be the most important to measure?
2. How is the communication performance of teams typically assessed in a JADC2 or similar context?
3. How is performance typically measured by instructors in a JADC2 or similar context?
4. How is performance typically measured within JADC2 (or similar) training systems (if it is measured at all)?
5. What human performance assessments would be valuable to be "baked in" to the STE?
6. Are there other aspects of general team performance that should be measured? If yes, please explain how they should be measured.
7. Do you have any remaining ideas, comments, or questions about data collection in the STE? If so, please record them here.
8. Do you have any remaining ideas, comments, or questions about performance evaluation in the STE? If so, please record them here.

Table 4 summarizes the themes that appeared in the respondents' answers to these questions.

**Table 4: Analysis of Open-ended Responses to Data Capture and Performance Evaluation Questions**

| Question | Analysis |
|---|---|
| What kind of performance variables do you think would be the most important to measure? | A number of responses to this item focused on the speed with which the team could select a COA (n=5) and effectiveness of the COA (n=6). Within the "speed" theme, respondents noted both "speed" (n=1) and the more nuanced "timeliness of orders relative to execution" (n=4). The "accuracy" theme included "accuracy" (n=2); "accuracy of plan of execution" (n=2); "accuracy of recommendations on COA" (n=1); and "effectiveness" (n=1). One response noted that the ease with which the COA could be interpreted/understood should be assessed.<br><br>A second broad theme that emerged from the responses focused on communications within the team. Three respondents contributed six measures related to communications. The measures were:<br><br>1. The ability for communications to make it from sensor to shooter/decision maker (n=2)<br>2. A record of who communicates with whom (n=1)<br>3. The quality/operational value of the communication (n=1)<br>4. The volume/number of communications among the JADC2 nodes (n=1)<br>5. The timeliness of the communications (n=1)<br>6. An assessment of whether critical communications get lost in the noise of lesser communications (n=1)<br><br>Closely related to measures of communication are measures of data flow that the respondents suggested. Data flow measures included the amount of information (bandwidth) passing through the system (n=2), measures of capacity and efficiency (n=1), and the time latency of information delivery (n=1).<br><br>Respondents also mentioned a handful of measures that did not fall within these three global themes (n=1, in all cases):<br><br>- Level of uncertainty<br>- Readiness<br>- Mission<br>- Maintenance<br>- Human Machine Interface validation |

| Question | Analysis |
|---|---|
| How is the communication performance of teams typically assessed in a JADC2 or similar context? | Responses to this query generally identified qualities of communication that should be evaluated or methods used to conduct the evaluation (n=1). Within the "qualities" theme, five respondents offered eight recommendations for dimensions of assessment:<br><br>1. Quality (n=1)<br>2. Appropriateness of information transmitted (n=2)<br>3. Timeliness of communication (n=2)<br>4. Correctness (n=2)<br>5. Completeness (n=2)<br>6. Appropriateness of the target of communication (n=1)<br>7. Frequency (number) of communication (n=1)<br>8. Conciseness (n=1)<br><br>Only one response fell into the methods category: "Expert Observation."<br><br>One other response theme seemed to span these categories: "Flow of information" (n=2).<br><br>Two respondents stated that they did not know enough to offer a response. |
| How is performance typically measured by instructors in a JADC2 or similar context? | The most frequent response was that the respondent did not know enough to offer a response (n=5). However, another popular response was that team performance was inferred from outcomes (n=4). A single respondent amplified this theme by noting that instructors evaluate the correctness, completeness, and timeliness of performance. |
| How is performance typically measured within JADC2 (or similar) training systems (if it is measured at all)? | The most frequent response was that the respondent did not know enough to offer a response (n=3). Echoing a theme from the preceding item, two respondents emphasized the subjective comparison of observed vs. desired behaviors. One participant responded "STRIKFOR." |

| Question | Analysis |
|---|---|
| What human performance assessments would be valuable to be "baked in" to the STE? | Five broad themes were present in the responses: Time (n=3); Outcomes (n=2); Processes (n=3); Individual State (n=3); and what the research team labeled "Covariates" (n=2). Respondents wanted the STE to capture a ***timeline*** of assessments, decisions (n=2), and the time required to complete various tasks (n=1). The ***outcomes*** of interest were outcome quality (n=1) and assessment accuracy (n=1). Within the "***Processes***" theme, respondents suggested the number of communications made (n=1), a catalog of the times that participants went outside the chain of command or the established/formal communications pathway (n=1), and measures of information fusion/filtering (n=1). The ***individual states*** of interest included workload (n=1), level of frustration (n=1), and level of stress (n=1). The ***covariates*** were variations in task conditions and included the number of assessments required within the scenario (n=1) and the relative difficulty imposed by data quality or variance within the scenario (n=1).<br><br>Two other responses referred to the management of the team (n=1) or, more generally, personnel (n=1). One participant did not know enough to offer a response. |
| Are there other aspects of general team performance that should be measured? If yes, please explain how they should be measured. | Respondents noted that it would be good to capture measures of team interaction/communication, such as a team dimensional training score (n=1), the number of interactions (n=2), and the quality of those interactions (n=1).<br><br>The respondents also suggested various measures of individual differences that might be valuable. This included the level of situation awareness that an individual had (n=1), the individual's level of fatigue (n=1), and the sense of overload that the individual was experiencing (n=1). One respondent rolled these individual measures up to the team level and noted that it would be important to characterize team state.<br><br>One respondent emphasized the importance of capturing domain-dependent outcomes such as speed and accuracy. Another said that he did not know enough to offer an opinion. |
| Do you have any remaining ideas, comments, or questions about data collection in the STE? If so, please record them here. | Respondents did not provide any new responses. |

| Question | Analysis |
|---|---|
| Do you have any remaining ideas, comments, or questions about performance evaluation in the STE? If so, please record them here. | Most respondents did not have anything additional to offer (n=7). Other respondents suggested (n=1 in all cases): <br>• The modes of communication used within the team might provide useful insights. <br>• The performance of agents should be assessed using human-in-the-loop assessment. <br>• Researchers should include agents/automation within teams. |

## 5.2   Assessment of Discrete Features

Sonalysts' experience in modeling, simulation, and training allowed us to predict a range of data collection and performance assessment features that we thought we should include in the STE. Within the survey, we asked respondents to indicate the degree to which they agreed or disagreed with our hypotheses about which specific features would be important within the STE. Pie charts and descriptive statistics for each of the domain-related features are shown in Appendix C.

As we did earlier, in this section we draw attention to items that surpass our threshold by having standard deviations greater than or equal to 1.0. In this case, only one item exceeded our standard for "significant disagreement." The item that had a broad range of responses was "The STE should have a feature to provide operators with real-time feedback on their performance." Responses on this item were split between "disagree," "agree," and "strongly agree," with the highest number of responses being "agree" (see Figure 5; standard deviation = 1.24). Yet 4 of 12 respondents disagreed that this should be a feature, making this a good topic for further discussion.

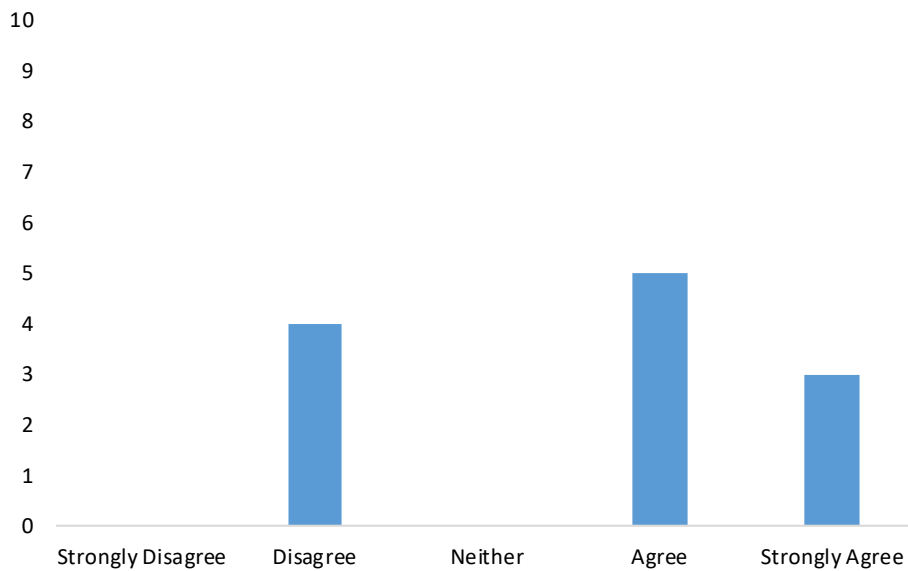

**Figure 5. Agreement ratings for "The STE should have a feature to provide operators with real-time feedback on their performance."**

## 5.3 Summary of Findings

The SMEs who participated in the survey indicated that it would be important to assess the communication within the team(s) participating in STE-based studies. The goal was to assess the ability of critical information to make it from the various sensors to the key decision-makers, and the efficiency with which that occurred. Responses to the open-ended survey items indicated that the SMEs largely focused on three key areas:

1. Was the communication directed to the correct teammate?
2. Did the communication include useful information?
3. Was the communication sent at the right time?

In general, the SMEs thought that it would be useful to monitor and assess the pattern of communication among team members to assess factors like the total amount of communications traffic and the specific patterns of communication (*e.g.,* who communicates with whom or the presence of road blocks that interfere with the flow of information). Beyond a quantitative assessment of team communication, the SMEs also felt that it would be important to assess its quality. They emphasized assessing the quality/operational value of the communication and the "signal to noise ratio" in the communication stream (*i.e.,* to what extent do critical communications get lost in the noise of lesser communications). The third dimension of performance assessment for communications was the timing of those messages. Was information "pushed" when it was appropriate or was it often "pulled" from lower echelons? Did critical information reach decision-makers in a timely fashion? Did certain interventions improve the speed of information flow? The SMEs noted that existing assessment protocols, such as Team Dimensional Training, might facilitate the assessment of communications within the STE.

Another data collection and performance evaluation focus for the SMEs was the assessment of the COAs that were selected and promulgated. A clear focus here was on the accuracy/appropriateness of the selected COA. The SMEs wanted to know whether participants chose the right/best COA and how quickly they did so. Beyond that, the SMEs wanted to get insight into whether the COA was effectively promulgated. They identified questions such as:

- Was the COA easily interpreted by the execution team?
- Was the COA promulgated quickly enough to enable timely execution?

Once the plan was selected and promulgated, the SMEs were interested in how well it was executed and how well it achieved the operational objective.

In addition to these two primary themes, the SMEs identified a number of other useful attributes that could be measured within the STE:

- Operator Workload
- Fatigue
- Level of frustration
- Level of SA
- Monitoring performance
- Speed and accuracy of task performance
- Overall outcome quality

In considering the methods that are typically used to assess performance, the SMEs noted that, for better or worse, these assessments largely focused on outcomes and were conducted on the basis of experts observing teams in action and providing subjective comparisons between observed and desired performance.

That said, in addition to the measurement of outcome variables, an analysis of discrete features revealed general agreement with the measurement of process and team coordination variables. Respondents either agreed with, or were neutral to, the inclusion of the following features in the STE:

1. Performance evaluation in the STE should include task outcome variables (*i.e.,* if the desired outcome is achieved).
2. Performance evaluation in the STE should include task process variables (*i.e.,* WHAT actions are taken and HOW the actions are performed).
3. Performance evaluation in the STE should include how well the team members coordinate with one another.

## 5.4 Topics for Further Exploration

A topic that respondents appeared to have mixed responses to is the notion that the STE should provide operators with real-time feedback on performance. Since the majority of responses were positive toward this feature, but a third of responses were negative, it may be a good idea to use opportunities to interface with this stakeholder group to clarify concern with this feature. Identifying the perceived drawbacks associated with this feature can help us understand whether the problem is one of design (incorporating a feedback feature is not worth the time needed to program it) or of principle (feedback will impede research in the STE).

## 6 DISCUSSION OF COMMUNICATION FEATURES

Another set of STE features that we wanted to explore addressed communication among teammates. Sonalysts' research team wanted to get a sense for how teammates typically communicated in JADC2-like settings.

### 6.1 Analysis of Open-Ended Items

The survey included five items related to communication within the STE:

1. What specific communication modalities are most relevant in a JADC2 or similar operational environment (*e.g.*, face to face, radio, written messages, *etc.*)?
2. What sorts of interfaces are used to support communication?
3. What interfaces are used to support other team-related behaviors?
4. Do you have any remaining ideas, comments, or questions about interfaces used in JADC2 (or similar) environments? If so, please record them here.
5. Do you have any remaining ideas, comments, or questions about team communication features to include in the STE? If so, please record them here.

Table 5 summarizes the themes that appeared in the respondents' answers to these questions.

Table 5: Analysis of Open-ended Responses to Communications-Related Questions

| Question | Analysis |
|---|---|
| What specific communication modalities are most relevant in a JADC2 or similar operational environment (*e.g.*, face to face, radio, written messages, etc.)? | Responses to this prompt fell into four discrete categories: verbal/spoken, written, face-to-face, and system data. Six respondents identified variations of **verbal/spoken** communications: Verbal (n=1); verbal over radio frequencies (n=3); verbal over Voice Over Internet Protocol (VOIP) circuits (n=2); voice within Command and Control systems (n=1); Satellite Communications (SATCOMS; n=1); and phones (n=1). Seven participants identified four categories of **written** communication: written (n=1); chat (n=4); email (n=3); and tasking and status messages (n=2). Five respondents identified **face-to-face** communications, with one specifying that those communications could be through secure video teleconferencing. **System data** included parameter values (n=1), a common operating picture (n=1), and electronic data transfer (n=1). |
| What sorts of interfaces are used to support communication? | Respondents identified a range of interfaces that could support team communication:<br><br>- Computers/Internet (n=5)<br>  - Cloud (n=1)<br>  - Internet Protocol (n=1)<br>- Radio (n=3)<br>- System Displays/Graphical User Interfaces (GUIs) (n=2)<br>- In-Person (n=2)<br>- SATCOMS (n=1)<br>- Email (n=1)<br>- Voice within Command and Control Systems (n=1)<br>- Phone (n=1)<br>- Paper (n=1)<br>- Direct Brain Interfaces (n=1)<br>- xR Interfaces (n=1) |

| Question | Analysis |
|---|---|
| What interfaces are used to support other team-related behaviors? | The responses to this item were largely in keeping with the previous two prompts. Four respondents identified five interfaces related to voice communications: voice communications (n=2); internet protocol (n=1); SATCOMS (n=1); Command and Control systems (n=1); and phone (n=1). Similarly, four respondents identified four variants of written communication (n=1, in all cases): Written, email, paper, and chat/messenger. Three respondents also identified face-to-face communications as an important interface. Other "repeats" were system displays (n=1) and Extended Reality interfaces (n=1).<br><br>Respondents also identified some novel interfaces:<br>- Computer/Internet (n=2)<br>- Collaboration Tools (n=1)<br>- Charts (n=1)<br>- Maps (n=1)<br>- Battlespace tracking (n=1) |
| Do you have any remaining ideas, comments, or questions about interfaces used in JADC2 (or similar) environments? If so, please record them here. | Respondents did not provide any new material in response to this question. |
| Do you have any remaining ideas, comments, or questions about team communication features to include in the STE? If so, please record them here. | Participants offered three additional thoughts in response to this item (n=1, in all cases):<br>- "We operate in a digital environment and speed is of the essence."<br>- "In today's world…..overload of information is a definite possibility. I find that when working tactical operations the best methods are visual, spoken comms and direct digital action (*i.e.*, VAB that authorizes engagement, *etc.*). CHAT, EMAIL and paper become a distraction."<br>- "Not the route I was thinking, but email and paper-based (letters, directives, *etc.*) are still necessary tools." |

## 6.2 Assessment of Discrete Features

As we have done with other sets of features, the research team presented respondents with a variety of communication-related features, and we asked respondents to indicate the degree to which they agreed or disagreed with our hypotheses about which specific features would be important within the STE. Pie charts and descriptive statistics for each of the domain-related features are shown in Appendix D.

More than half of the items in this section (three out of five) satisfied our "standard deviation greater than 1" standard. All three pertained to the medium of communication within the STE. One feature with some disagreement about its value in the STE was "The STE must enable video-based communication among teammates." This feature had a standard deviation of 1.08. While most responses fell into "agree" or "strongly agree" categories, one third of respondents disagreed, or were neutral to including the feature.

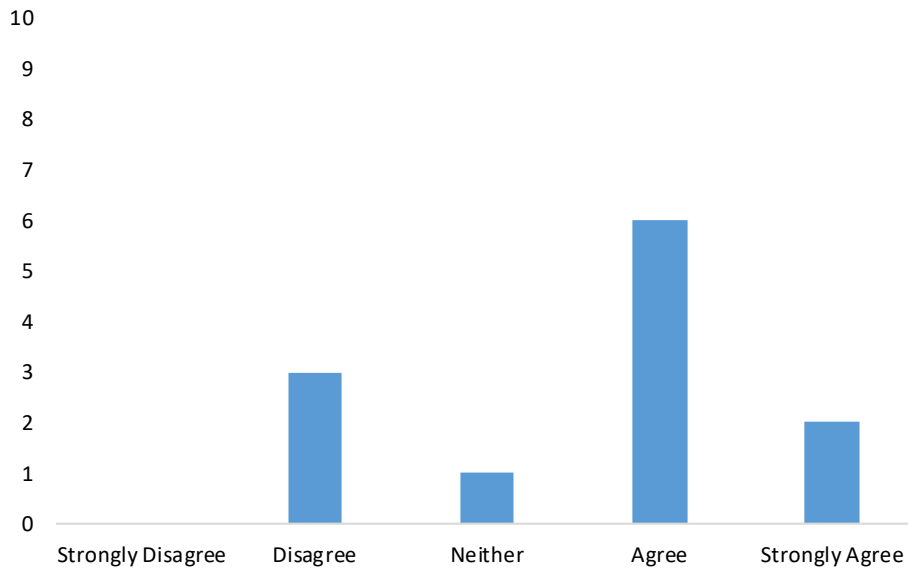

**Figure 6. Agreement ratings for "The STE must enable video-based communication among teammates."**

Another feature with a lack of agreement was "The STE must enable paper-based communication among teammates." This item was more distributed over "strongly disagree" and "disagree" options, although as with the video-based communication feature, the most frequent response was "agree" (see Figure 7; standard deviation = 1.04).

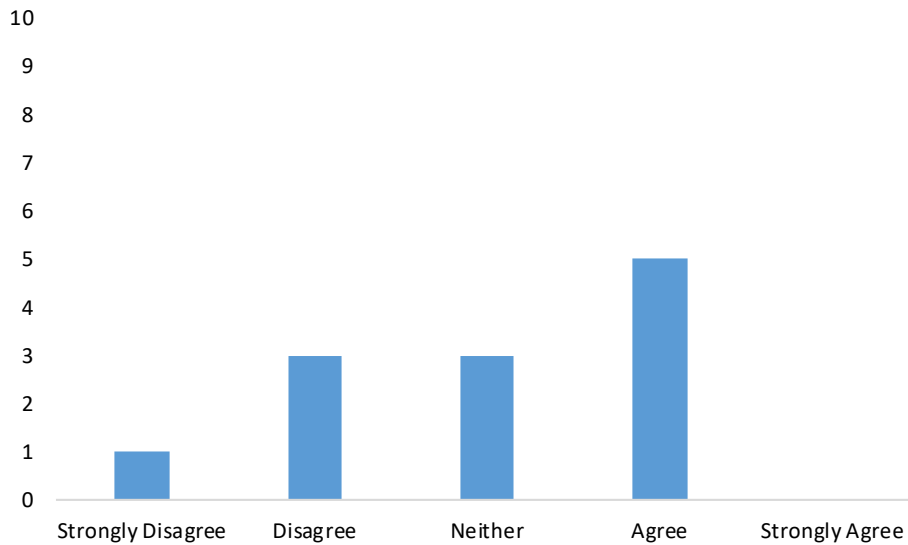

**Figure 7. Agreement ratings for "The STE must enable paper-based communication among teammates."**

The feature "The STE must enable spoken communication among teammates" also had mixed responses (standard deviation = 1.15). Most were concentrated at the higher end of the scale with strong agreement. However, one respondent strongly disagreed that this feature should be a part of the STE (Figure 8).

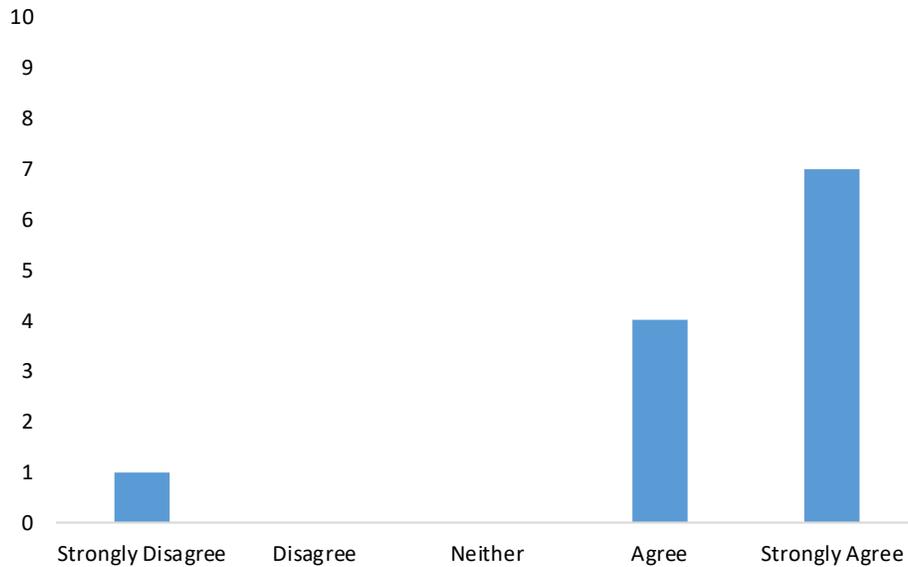

Figure 8. Agreement ratings for "The STE must enable spoken communication among teammates."

## 6.3 Summary of Findings

The SME respondents identified four communication modalities and the associated interface types. As we develop the STE, we should consider incorporating these modalities and types.

The first modality was simply face-to-face communication. This might involve walking across the room or down the hall to speak with someone. Alternatively, the face-to-face interaction can be supported with video-based teleconferencing support (*i.e.*, Zoom, Teams, or something similar). However, supporting video communication was somewhat controversial. Identifying the best way to incorporate this into the STE may require some additional discussion with SMEs.

Another common communication modality might be listed as "aural." This is spoken (not visual) communication that takes place when individuals are geographically separated. This can take place using traditional radio links, VoIP connections, SATCOMs, or other dedicated circuits within a given Command and Control system.

The SMEs also noted that written communication was quite common. This includes real-time text-based chat messages, asynchronous email exchanges, and formal tasking and status messages. The written messages can be delivered electronically or on paper. Consistent with this, respondents mostly agreed with (or were at least neutral to) the idea that the STE should enable (1) email communications and (2) chat-based communications among teammates.

The fourth modality that respondents mentioned was "system data." This includes using the combat system to share parameter values, a common operating picture, and other similar planning artifacts.

When the discrete items are considered, a clear priority emerged:

1. Spoken communications ($\bar{x}$ = 4.33)
2. Chat-based communications ($\bar{x}$ = 4.25)

3. Email communications ($\bar{x}$ = 4.00)
4. Video-based communications ($\bar{x}$ = 3.58)
5. Paper-based communications ($\bar{x}$ = 3.00)

## 6.4 Topics for Further Exploration

The communication topic that generated the most disagreement was the incorporation of features related to various communication methods and interfaces. Video- and paper-enabled communication had the greatest controversy. This disagreement may be fertile ground for discussion with an opportunity to follow up with SMEs.

## 7 DISCUSSION OF AGENT FEATURES

The primary focus of Sonalysts' research plan is to explore various hypotheses regarding the use of training to improve the performance of Human-AI teams by enhancing the ability of humans to function as part of those teams. However, in the course of that research, it is likely that we will also explore features and/or capabilities of autonomous teammates that have a similar impact. For this reason, we wanted to give our SME respondents the opportunity to comment on the agent features/capabilities that are likely to be important as the agents team with humans in a JADC2 or similar context. This section explores those features.

### 7.1 Analysis of Open-Ended Items

The survey included four items related to autonomous teammates within the STE:

1. What autonomous systems technologies do you currently use?
2. What characteristics of an autonomous teammate would lead you to trust or distrust it? Why?
3. What characteristics of autonomous teammates do you think are important to include in a human-AI teaming testbed?
4. Do you have any remaining ideas, comments, or questions about autonomous teammates in the STE? If so, please record them here.

Table 6 summarizes the themes that appeared in the respondents' answers to these questions.

**Table 6: Analysis of Open-ended Responses to Autonomous Agent Questions**

| Question | Analysis |
|---|---|
| What autonomous systems technologies do you currently use? | Most respondents (n=5) indicated that they did not use autonomous systems. Five respondents did identify six systems:<br><br>• Recommender Systems (n=2)<br>• Automotive Safety and Navigation Systems (n=2)<br>• Game-based Autonomy (n=1)<br>• Data Collection and Display Tools (n=1)<br>• Low-level autonomous systems (n=1)<br>• Basic shipboard automation systems (n=1) |

| Question | Analysis |
|---|---|
| What characteristics of an autonomous teammate would lead you to trust or distrust it? Why? | Six themes emerged in the analysis of features said to promote trust in autonomous teammates. The most cited theme was having an understanding of the agent's decision logic (n=3) and related items such as data transparency (n=2); understanding of how the agent was trained/developed (n=1); availability of explanations (n=1); and alignment with human expectations/experiences (n=1). Familiarity with the agent (n=1) and a history of good performance (n=2) were also cited as promoting trust. Other trust-enhancing features included:<br><br>• Ability to signal when the problem is "out of bounds" (n=2)<br>• Ability to smoothly reintegrate a human for out of bounds problems (n=1)<br>• Overlay or feature to view "raw" sensor data (n=1)<br>• Actual assistance in the task (*i.e.*, is the information helpful) (n=1)<br><br>Features that were said to damage trust were:<br><br>• Actions that seem contraindicated or erratic (n=2)<br>• Surprise (n=1)<br>• Actions taken without "human interface" (n=1)<br>• Faulty information (n=1)<br>• Biases (n=1) |

| Question | Analysis |
|---|---|
| What characteristics of autonomous teammates do you think are important to include in a human-AI teaming testbed? | The research team was able to group responses to this prompt in four global themes: Supporting human understanding, acting as good teammates, having good performance, and processing/reacting to dense data streams.<br><br>To **support human understanding**, respondents wanted the agents to offer explanations (n=2), have good communication skills (n=1), and support voice interaction (n=1).<br><br>Acting as **good teammates** included the ability to signal when the problem is "out of bounds" (n=1); the ability to reintegrate humans smoothly for out-of-bounds problems (n=1); the ability to support human-in-the-loop decision-making (n=1); the ability to manifest a sense of urgency (n=1); and a desire to hold agents to the same standards as humans (n=1).<br><br>Four ideas were grouped within the **having good performance** theme: a good performance track record (n=1); the ability to retrain rapidly from non-optimal responses (n=1); behaviors that are both conventional and novel, but correct for the situation (n=1); and the ability to improve on the *status quo* (n=1).<br><br>Examples of the ability to **process and react to dense data streams** included deciphering and analyzing multiple data sources simultaneously at a rapid rate (n=1); identifying important data (n=1); providing timely alerts (n=1); limiting both false alarms (n=1) and misses (n=1); and avoiding being led astray with faulty/unreliable data (n=1). |

| Question | Analysis |
|---|---|
| Do you have any remaining ideas, comments, or questions about autonomous teammates in the STE? If so, please record them here. | One set of responses reflected a general skepticism regarding the value of autonomous teammates:<br><br>• "I am not a fan of 'on the loop'……and will always differ [sic] to 'in the loop'. Decisions of weapons, navigation and/or matters that could lead to injury and/or death need to always have a 'human' component." (n=1)<br>• "I don't think that autonomy needs to replace all human functions." (n=1)<br>• "I don't think it should be constrained by chain of command issues as long as it isn't in full automatic mode." (n=1)<br>• "An Autonomous Teammate will never be allowed to replace a human in the fire control loop due to legal statutes." (n=1)<br><br>Other responses identified additional desirable features of autonomous teammates:<br><br>• "Autonomous teammates must monitor the evolving situation and respond proactively." (n=1)<br>• "Focus on basic tasks first; sensor correlation and fusion, mission/task assignment, status, progress reports, etc." (n=1)<br>• "Behave as much like a human teammate as possible." (n=1)<br>• "Provide explanations/qualifiers, as appropriate." (n=1) |

## 7.2 Assessment of Discrete Features

Our review of the relevant literature identified various agent features/capabilities that were likely to be useful in autonomous teammates. As in other portions of the survey, the research team presented respondents with several of these features and asked respondents to indicate the degree to which they agreed or disagreed with our assertions regarding which specific features would be important within the STE. Pie charts and descriptive statistics for each of the domain-related features are shown in Appendix E.

Of the 11 agent features included in the survey, only 1 item has a standard deviation greater than or equal to 1.0. The feature "Within the STE, the roles that humans fill and the roles that autonomous teammates fill should be interchangeable" had a standard deviation of 1.09. The most frequent response was to disagree that this feature should be part of the STE. However, there was at least one response in favor of every point on the agreement scale (see Figure 9).

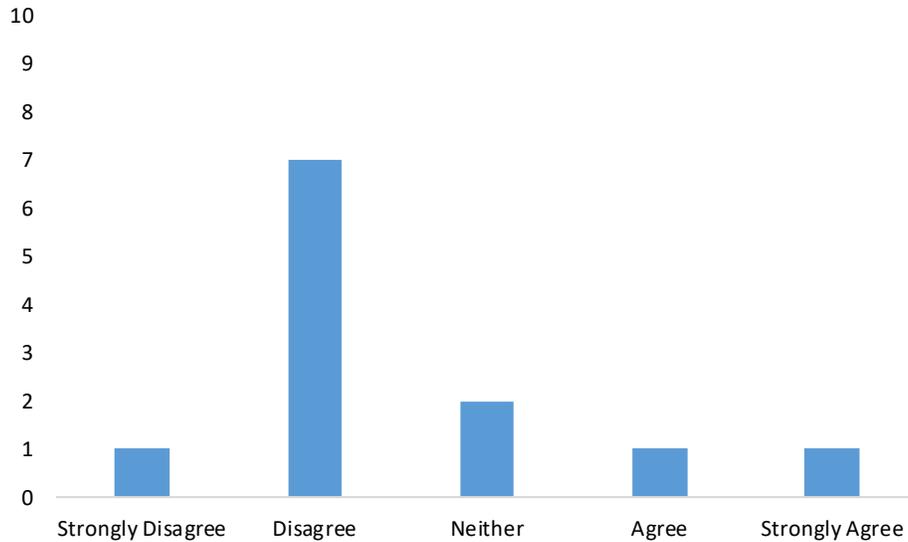

**Figure 9. Agreement ratings for "Within the STE, the roles that humans fill and the roles that autonomous teammates fill should be interchangeable".**

## 7.3 Summary of Findings

It is worth noting that our SMEs had little or no experience with autonomous agents, and certainly none that were approaching the maturity to be considered "teammates." Most of their experience came from "low-level autonomous systems" such as commercial products like automotive safety and navigation systems, non-player characters within computer games, or various recommender systems. In a military setting, the "agents" took the form of system automation tools and data collection and display tools.

The lack of familiarity was paired with significant skepticism regarding the ability of agents to function as reliable teammates. For example, one respondent noted that he/she was, "not a fan of 'on the loop'……and will always differ [sic] to 'in the loop'." One respondent suggested their trust in the agent would be harmed by the agent taking action without a human's permission, stating that "Decisions of weapons, navigation and/or matters that could lead to injury and/or death need to always have a 'human' component." Another respondent questioned whether current laws would even allow an autonomous teammate to be involved in weapons employment.

The respondents were able to describe features that would lead them to trust or distrust the agent. These are summarized in Table 7.

**Table 7: Features that Promote or Harm Agent Credibility**

| Features that Promote Credibility/Trust | Features that Harm Credibility/Trust |
|---|---|
| The ability to understand the decision logic of the agent. This includes:<br><br>• Understanding of how the agent was trained/developed<br>• The transparency of the data used to train the agent<br>• The ability of the agent to provide meaningful explanations<br>• The alignment of the agent's performance with the human's expectations/experiences | Taking actions that surprise/confuse the human members of the team. |
| The agent's ability to provide an indication of its confidence in its performance. This includes:<br><br>• Ability to signal when the problem is "out of bounds"<br>• Ability to reintegrate human smoothly for out of bounds problems | The agent is "led astray" by faulty data. |
| A reasonable period of time working with an agent and seeing that it has a good performance track record. | |

An analysis of discrete features echoed some of the responses in the open-ended format questions. The items for which there was the most agreement were:

1. Autonomous teammates within the STE must be able to acknowledge inputs from other teammates ($\bar{x} = 4.25$).
2. Autonomous teammates within the STE must be able to temper trust by actively monitoring for conditions in which failure was likely and notifying teammates that they should not rely on the automation in such situations in the future ($\bar{x} = 4.25$).
3. Autonomous teammates within the STE must be able to adhere to a chain of command ($\bar{x} = 4.25$).
4. Autonomous teammates within the STE must be able to accurately monitor and report their status ($\bar{x} = 4.25$).
5. Autonomous teammates within the STE must be able to accept guidance from teammates (human or non-human) as conditions and priorities change ($\bar{x} = 4.50$).
6. Autonomous teammates within the STE must be able to notify teammates of on-going progress, state changes, etc. ($\bar{x} = 4.67$).

Consistent with some research in this area, our SME respondents wanted to hold autonomous agents to a high-level of performance, stating that the agents should be held to the same standards as humans, but then leavening on additional requirements. Desirable features listed by the respondents included:

- The ability to ingest and analyze multiple data sources simultaneously at a rapid rate.
- The ability to identify the importance of individual data items and prioritize them appropriately.

- The ability to provide human team members with critical alerts.
- The ability to offer meaningful explanations of actions/recommendations.
- The ability to interact with human teammates using normal speech.
- The ability to embody a sense of urgency.
- The ability to respond to an evolving situation pro-actively.
- The ability to learn from non-optimal responses.

### 7.4 Topics for Further Exploration

The most controversial feature for the STE regarding autonomous agents is the notion that the roles of humans and agents should be interchangeable. This is consistent with some of the long-form answers we received on this topic. For instance, the notion that agents should not perform any tasks without the human's permission assumes that the roles that humans and agents occupy must at least differ to the extent they are able to perform certain actions. This is another area of potentially fruitful discussion among SMEs in further interface with this stakeholder community.

## 8 NEXT STEPS

Based on the survey responses and analysis, we do not intend to conduct a follow-up Targeted Ideation Development Event (TIDE) with SMEs. Many of the responses to questions related to JADC2 tasking and teaming structure suggested that much is still unknown about JADC2 and how it will be implemented. It is not likely that a TIDE will be able to move the group toward consensus until more clarity on the JADC2 concept of operations exists.

We will consider other engagement approaches in the context of the results of the "researcher" survey. For example, the results of that survey may suggest the need to complete a TIDE, focus group, or some other form of follow-up engagement. At that time, we may want to consolidate the SME and researcher portions of the stakeholder community.

This Page Intentionally Blank

*Appendix A*

*SME Knowledge Elicitation Survey Responses*

In this appendix, we list the respondents' answers to each question. The following conventions are used:

- Bold and italicized text indicates a survey question.
- Text in quotations indicates a response to the question.
- Line breaks are used to separate respondents' answers.
- Answers that were "non-responsive" (*e.g.*, "I don't know") are not included.

***How are JADC2 teams structured?***

"I don't know, but would imagine similar to existing C2 structures."

"Based on current US Combined Air Operations Center (CAOC) Structure…All components Land (Battlefield Coordination Detachment- BCD); Maritime (Naval and Amphibious Liaison Element-NALE); Special Ops (Special Operations Liaison Element-SOLE); Cyber Elements; and Coalition Elements."

"Joint Staff, Army, Navy, Air Force."

"In my last command (N2C2), the JADC2 teams were structured by domain (air, land, maritime, cyber, missile & space) and by function (COP, missile defense, emergency action controller, help desk, intel, messages, and current operations)."

"By component. Within components there are working groups with different areas of responsibility."

"I don't have good knowledge of JADC2 as a concept. C2 is normally accomplished by having a primary command role with supporting staff organized by competency. Often it is centered on the types of platforms involved (air vs surface vs subsurface as an example)."

"Lead service determines structure…..example: in 7th Flt, AF Wing was the lead for AD, therefore they were the theater ADC and each component had a rep."

"JADC2 teams are structured based on the specific task assigned and the resources available at the time. The structure for a mission primarily sea-undersea focused would be different from the structure for a sea-land or an air-land focused structure. Consider that the command and control structure for warfighting in Iraq would be different from the ASW command and control structure for the North Atlantic."

"Currently or in the future? There are domain teams within the staff that execute within each domain. For example, the Joint Land Component Commander and his team handle the land aspects of the all-domain theater. I am not sure that best serves the all-domain gains desired by the combatant commanders."

"They likely include a communications staff, multiple intelligence liaisons, representatives for each community of each branch involved as SMEs/advisors, and a commander."

***Generically, what types of tasks are performed in a JADC2 environment?***

"Sensor fusion, tracking, targeting, Maritime Domain Awareness? (MDA) (or whatever the AF/Army equivalent of that is?)."

"I would expect: evaluation of commander's intent, targeting cycle, execution planning, collateral damage assessments."

"In current operational constructs, Planning Guidance and Objectives are Promulgated to the Component commands (Land, Sea, Air, Maybe Space and Cyber)…To me JACDC2 implies a capability within the Joint COCOM Staff…Currently a COCOM staff is really man, trained and equipped to conduct operational level planning and monitor…The envisioned JADC2 functions of any sensor, any decision maker would require new capabilities at the Joint Force Commander to conduct operations further down the operational/Tactical spectrum…My understanding is that this level of thought has only been provided cursory thought. All of the efforts so far in Confluence Project have revolved around making individual capabilities talk to each other. The US Military is a long way from the JADC2 Vision of C2 in the Joint warfighting Concept."

"Connecting all communication (Joint, not just one service) to be universal. For example, an observer on a different platform can provide targeting information to a different platform. Communications as a meshed network essentially - a cloud concept."

"Battle management of the assets, coordination and information gathering between upper and lower echelons, maintaining SA and ensuring the COP (respective to your domain) is accurate and timely for senior decision makers."

"Coordination, planning, de-confliction."

"A C2 environment will facilitate planning, communications of plans, status reporting and plan adjustments. It also supports engagement decision making."

"Mission and asset management, priorities and de-confliction."

"The tasks would be a function of the assigned mission. A JADC2 environment is the operating and warfighting environment in which the joint force finds itself. In a general sense, the joint force must be capable of conducting all possible tasks."

"Sense, decide and act. Plan, deploy/maneuver, react. Orders, plans, and reports to higher authority."

"I would say the primary task is information fusion - collection of real-time data, analytics, and intelligence to rapidly build the battlespace picture for decision-makers to effectively execute command authority."

"F2T2EA. But there needs to be more. I will attach an unfinished white paper that explains more."

*Is the OODA loop a good metaphor for how JADC2 operations are conducted? If not, what is a better metaphor?*

"Possibly."

"Probably yes."

"Not quite sure the OODA Loop is quite sophisticated enough for the complexities of JADC2…may be a starting point, but the fact that there are multiple cycles that are promulgated within US Joint, service, alliance, and coalition operations…Maybe JADC2 will harmonize but I doubt it. Since JADC2 is envisioned (Keep in mind a vision without resources is a Hallucination!) to be a joint level any sensor; any shooter capability… The 6 phase Joint targeting could have value see https://apps.dtic.mil/sti/pdfs/ADA403414.pdf."

"Yes, as the desire is to react inside the enemy's decision space through all connected communications."

"I suspect that in phase II/III operations that is a good characterization. In phase 0/I operations (shape/deter) there are good I/W indicators that lead times are sufficient where the urgency of the OODA loop is not exercised."

"Yes."

"OODA is a good metaphor if extended to the theater level. Typically I've seen the term applied at the unit level, but I think similar principles apply at the theater level."

"The OODA loop is not a useful metaphor for operations where JADC2 is employed. OODA is potentially useful for tactical level of warfare. It is not useful for the operational or strategic levels of warfare where JADC2 provides the most value to the commander. As more useful metaphor is some sort of networked depiction with links and nodes, where the nodes are the people, machines, and organizations that are entities executing the missions and tasks. The links are the relationships and information flows among the nodes."

"The OODA loop is as good as any."

"I feel that the OODA loop is almost inescapable when it comes to military operations. It is still very relevant and really what we're doing now is scaling the speed at which the "actions" occur and compressing the "observe-orient-decide" part."

"Again, my white paper will address this. OODA is a good start, but many use it as a dashboard crutch. Boyd did not mean it as such. It has uses on many levels."

***Are there hierarchical relationships in JADC2 teams (i.e., a clear command hierarchy) or are team members typically more parallel in authority? Does this change based on tasking?***

"My initial thought is there probably has to be in order to assign ownership and establish a formal comms path. This is usually the largest problem with joint ops that I've seen. Maybe? But I think you want at least the "backbone" to remain the same."

"I would expect a clear hierarchy, but I'm not positive."

"I am not familiar with JADC2."

"JADC2 will require changes to current force structures and doctrine. Current planning Doctrine is "Centralized Planning/ De-Centralized Execution". JADC2 Concepts (really thin at this time) could drive Planning and C2 at the COCOM level Centralized Planning AND Centralized execution). Current Doctrine: After COCOM guidance is received…the components enter into their own service-based Planning processes (done with coordination between the services. There are some nuances to this process based on emerging domains like Cyber and Space."

"Not sure but one must have a system that permits autonomy. Having decisions being required by higher authority results in the enemy getting inside you OODA loop as Ukraine is demonstrating with Russia hierarchal system. China is not much different."

"In my experience, my domain and functional teams were in parallel authority reporting to myself as the director."

"Yes there are hierarchical relationships but a lot of the collaboration is more matrixed."

"I believe it is hierarchal but below the top tier you have parallel authority in different domains."

"Always a C2 structure in place…..same as any other military operational command."

"There are absolutely hierarchical relationships - someone is in command; someone has to make resource allocation decisions among the various commands, and so on. Depending on the tasking, there will likely be some degree of parallel authorities."

"I don't think anyone knows where this needs to go. There is currently a clear structure from the combatant commander on down."

"I don't know specifically, but I would probably set a singular commander with a flat organization below comprising of the team members - ultimately this gives the commander the opportunity to weigh all information appropriately from each team member; no team members' thoughts or inputs are overridden by senior members within the team."

***Do you have any remaining ideas, comments, or questions about JADC2 operations? If so, please record them here.***

"Jim, There are myriad of things required to change, mature, or be developed from scratch that would be required to operationalize the JADC2 concept…we welcome any chance to noodle some of the deep thoughts we have on this tropic. [sic]"

"How will it work in a A2/DA environment? If you don't have access to the cloud, you are in deep trouble if you are relying upon it."

"No additional comments, except to ensure that the survey reflects the degree of sophistication necessary for JADC2."

"I think you have to be well aware of why JADC2 is important and drive the organizational structure to optimize against that. It will also identify the weak points and needs for improvement in kill chain processing necessary to succeed."

"I'm not very familiar with them and wikipedia and the internet are fairly scarce on information (appropriately so). So any 101 breakdown would be helpful."

"See the white paper I'll send on JLOCC."

***What is a good analogous task domain for the STE to incorporate (i.e., a domain that is analogous to the JADC2 task domain without being classified)?***

"Mountain/Wilderness/Open Ocean Search and Rescue?"

"What about some sort of paintball, capture the flag type of team competition that incorporates a command element with field forces executing tasks."

"Integrated Air and Missile Defense (IAMD) is probably the best domain…1) The new Army IAMD C2 system (IBCS) has been designated as the Army entre into the JADC2 System 2) IAMD is inherently joint in execution due to the doctrinal set up that all IAMD elements regardless of Service answer to the Area Air Defense Commander (AKA the JFACC)."

"I'm not familiar with the designs of an STE but I saw that the Uber Model is a good model (https://sgp.fas.org/crs/natsec/IF11493.pdf)."

"Maintaining SA of your domain to include processing and analyzing data that determines if action is required in a timely fashion."

"The NIMS Incident Command Structure (ICS)."

"This question is difficult to answer because one needs to specify a "level of scale" to which the question applies. If you are asking about tasks at the command or task group level, then the UJTL is a pretty good articulation of tasks that will have to be conducted by JADC2. If you are asking about tasks that individual watchstanders/operators or individual machines must accomplish, that is a much different question and would need to benefit from some degree of analysis at that level."

"9/11/2001."

***What qualities would that domain need to possess?***

"Different orgs, different sensors, overwhelming data."

"Needs to have analogous tasks for each level of decision that has be conducted in the classified version. It also needs to consider bandwidth requirements of the classified variant to ensure too much information isn't being provided/demanded."

"Plan Air and Missile Defense, Fire Control for AMD assets Any Sensor, any shooter, Sensor Management, Maintenance Management, Communications Integrated Fire Control Network (IFCN) Management, Supply Management, Personnel Management."

"Unlike UBER, war has the dynamic of your opponents desire to crush you and your advantage. How is it hardened, how do you work with A2/AD. How does it deal with non-continuous comms, e.g. a submarine."

"High degree of experience and understanding of the direct and indirect forcing functions within that domain environment."

"Structured, scalable, repeatable. A relatively loose framework that binds together over a rapidly evolving scenario - and which leaves no ambiguity about where and who makes decisions."

"See #6 above. Without some articulation of what is considered a "domain", providing qualities is difficult."

"I don't understand the question. Multiple, simultaneous axes of action and decision-making all rolled up into a response commander."

***Do you have thoughts about the "game" that participants should be asked to play within the STE?***

"Some sort of paintball, capture the flag type of team competition that incorporates a command element with field forces executing tasks."

"BAE's Pioneer model has potential. Also, I saw an article about Palantir system in Ukraine that looks to be very robust (https://www.washingtonpost.com/opinions/2022/12/19/palantir-algorithm-data-ukraine-war/)."

"Yes. Gaming scenarios need to stress the domain watch standard in order to prioritize and multitask activities accordingly."

"Tasks are assigned and teams perform tasks with communications to other teams and they pull info from a supporting task. AI/ML might assist in assimilating information and providing recommendations in a semi-automated way."

"See #6 above. A wargame to explore execution of a mission in a JADC2 environment would be much different from a wargame to explore operator and machine functions and employment as part of a JADC2 staff."

"Time critical, decision consequence, losing and winning along each axis."

***Do you have any remaining ideas, comments, or questions about the task domain used in the STE? If so, please record them here.***

"I think the pressure to win should suffice for the level of implementation needed."

"Continual feedback from real time, could adversely affect the operator with continual updates and reworks."

"Rather than uncertainty use data availability as your concept to drive the knowns and unknowns."

"I think these all hit very well on what I was trying to get at in my earlier comments/questions."

***What kind of performance variables do you think would be the most important to measure?***

"Time latency of information vs legacy systems."

"Amount of information (bandwidth). Accuracy of plan execution. Timeliness of orders relative to execution."

"Speed, Accuracy, Effectiveness, Human Machine Interface validation."

"Ability for comms to make it from sensor to shooter/decision maker."

"Time, "correctness", coordination."

"Level of uncertainty, accuracy of recommendations on COA, ease of interpretation."

"Readiness, mission, maintenance."

"- who communicates with whom? - what is the quality/operational value of the communication? - what is the volume/number of communications among the JADC2 nodes? What is the timeliness of the communications? Do critical communications get lost in the noise of lesser communications?"

"Time, accuracy, maybe a talk aloud recording, collection of what is needed versus what is available. I guess it depends on the objectives of the study. If you are trying to tease out how various organizing concepts work within JADC2 you would collect indicators of capacity and efficiency."

"Communication effectiveness, the ability to correctly recognize and analyze the information available, completeness of the information profile - exhausting all methods of data collection available within the timespan, timeliness of decision-making."

"Both direct and proxy measurements. But more importantly, were we able to set conditions through action, purposeful inaction, perception management, and deception? And was our deception able to ACHIEVE something, and not just offer a false impression? We also need to create measurements for creating Haven (I'll send a description of Haven)."

**How is the communication performance of teams typically assessed in a JADC2 or similar context?**

"Correctness, Timeliness, Completeness."

"Mission Dependent, but purely related to the transmission of the proper information to the proper Echelon, tin a timely manner to meet mission requirements…How much bandwidth is required to ensure mission success at Echelon is unknown to us at this time."

"How informed the final decision is."

"Through observation of experts."

"Flow of information."

"- numbers of communications without any sense of value."

"The flow of orders and reports of activity."

"Conciseness, accuracy, knowing what is most important at the time: don't fill the air with useless information. Be the expert in your area and provide only the information needed to make decisions. In general, don't interrupt other reports… unless your information is of primary concern now."

"We live in a world of overwhelming choice, less-than-attentive individuals, poor writers and speakers, and less-than-5x5 comms. To assess comms performance would be the number one priority, since comms make the difference among burning platforms, going winchester, or over-stimulating (aka escalation). We need to have just enough skin in the game without over-committing, and comms should be evaluated on whether that (as a goal) has been achieved."

**How is performance typically measured by instructors in a JADC2 or similar context?**

"Correctness, Timeliness, Completeness. I would hope teams are assessed on the things that matter in Q12."

"For JADC2 Unknown…not currently a system of record…hence no school. Any experimental operators will be provided OJT."

"End result."

"Perceived objectives goals versus actual."

"Depends on the grading factor and/or metrics being used."

"Mostly outcome measures. Decision making omissions."

"Overall success of the mission. Other grading is subordinate to that and can be nuanced, likely."

"No current knowledge, but I'm guessing it's purely tactical."

**How is performance typically measured within JADC2 (or similar) training systems (if it is measured at all)?**

"I would hope this is the same answer as Q13."

"I'm not familiar with other training systems."

"Currently doesn't exist in an institutional environment, only in the R&D/S&T realm."

"End result."

"By subjective assessment of goals vs achievements."

"Believe now STRIKFOR Training is now certifying Navy units for Joint operations."

"The "goods" and the "needs for improvement". There is normally a very senior exercise observer that imparts feedback to the commander."

"The training system should be accurately representative, but also have built in options - the command team should sometimes be able to make 1 of many decisions that result in varying levels of success. Basically, if they are always forced into 1 decision, it's not realistic."

**What human performance assessments would be valuable to be "baked in" to the STE?**

"Not sure it's a specific assessment, but if you had a "game" I think you would want to note every time participants went outside the CoC or established formal comms pathways."

"Time to complete various tasks in the tool."

"Frustration?"

"Decision cycle."

"Number of communications made, number of assessments, assessment accuracy, timeline of assessment and decision, relative difficulty imposed by data quality or variance."

"Workload, comprehension."

"- see the inputs for #11 above."

"Outcome measures."

"Stress management, personnel management, team management, information fusion/filtering."

"The ability to synthesize "setting conditions" with dealing with real-time, current operational issues. NUMBER ONE SKILLSET in the 21st C."

***Are there other aspects of general team performance that should be measured? If yes, please explain how should they be measured?***

"Fatigue/overload."

"This JADC2 function must also be compared to service or domain inputs for harmonization. Analysis would have different measures by different functions: Planning might include time measurement ; Logistics- maybe quantitative accuracy (i.e. was the amount of fuel required properly calculated)… All JADC2 Measures of Effectiveness would need to be developed for a myriad of different functions across the JADC2 System."

"Amount of coordination."

"Quantify team interactions throughout the approach to solving a problem… quantify the contribution of AI/ML."

"- performance of machines/computers/AI must be assessed in detail just like human performance. - performance of the human-machine team must be assessed in detail."

"SA and TDT."

"Mentioned above - but really just seeing how the team interacts generally. Is the communication direct, concise, non-ambiguous? Is everyone tense or are they calm? How do they react when something unexpected happens? General demeanor?"

"STE would need to have a way to calculate performance against the JADC2 effort being an HRO (Highly Reliable Organization), with the ability to operate in a non-linear, novel environment."

***Do you have any remaining ideas, comments, or questions about data collection in the STE? If so, please record them here.***

"Depends upon what you are trying to prove in the JADC2 context."

"See my white paper."

***What specific communication modalities are most relevant in a JADC2 or similar operational environment (e.g., face to face, radio, written messages, etc.)?***

"IP, SATCOMs, radio, written/manual inputs."

"Face to face verbal over IP verbal over RF short digital text like email type."

"Face to face by Secure VTC; Chat; Secure Email; Voice within Command and Control Systems."

"Chat."

"Face to face, chat, phone."

"Face to face, text, tasking and status messages, system data flow parameters."

"Yes to all the examples listed….and in addition chat, visual (i.e. COP)."

"- All of the modalities are relevant, in part, because each modality has strengths and weaknesses.  For example, email or written messages might well be best for routine communications, whereas face-to-face might well be best for critical communications and under conditions of high uncertainty."

"Electronic data transfer.  You have to greatly close the OODA loop.  Voice."

"Message traffic, radio, face-to-face in the JADC2 team."

"I'm sending another white paper titled Balance of Information. Please see that."

***What sorts of interfaces are used to support communication?***

"Pretty much all. Not sure there's a limit here."

"Internet RF path (not sure the frequency spectrum) In-person."

"Cloud – IP."

"Computer, phone."

"Air, software/network, paper, display."

"Headsets, GUIs/screens, visual."

"- standard interfaces  - direct-brain interface  - virtual/augmented reality."

"Radio and electronic data transfer."

"Satellite communications, other RF means (OTH, direct), computers and radio/communication equipment."

"Need to limit the bandwidth of over-tasked operators so that they focus. Let others deal with administrivia."

***What interfaces are used to support other team-related behaviors?***

"IP, SATCOMs, F2F, chat, radio, written/manual inputs."

"Computer, phone."

"- standard interfaces - virtual/augmented reality."

"Face to face and internal team voice comms."

"Collaboration tools, chat/messenger, charts, maps, battlespace tracking."

***Do you have any remaining ideas, comments, or questions about interfaces used in JADC2 (or similar) environments? If so, please record them here.***

"At this time there is very little information on operational JADC2 Concepts."

***Do you have any remaining ideas, comments, or questions about performance evaluation in the STE? If so, please record them here.***

"A human in the loop assessment of would be critical to measure Autonomous Decision Trust/Confidence that decision had a reasonable level of confidence."

"Communication mode is an attribute that may be the basis for comparisons related to effectiveness."

"- must include machines/computers as part of the JADC2 team at a variety of levels of scale."

"These are generally along the lines I was going, as well."

"Performance evaluation also should include WHY, especially in the context of setting conditions for future exploration, exploitation, or abandonment of conditions."

***Do you have any remaining ideas, comments, or questions about team communication features to include in the STE? If so, please record them here.***

"We operate in a digital environment and speed is of the essence."

"In today's world.....overload of information is a definite possibility. I find that when working tactical operations the best methods are visual, spoken comms and direct digital action (i.e. VAB that authorizes engagement, etc.). CHAT, EMAIL and paper become a distraction."

"Not the route I was thinking, but email and paper-based (letters, directives, etc.) are still necessary tools."

"See the Rules for Balance of Information in the Associative Age in the Balance of Information white paper (at the end of the paper) for a key inclusion of a comms/info evaluation."

***What autonomous systems technologies do you currently use?***

"None."

"SEII based autonomy."

"There may be some low-level autonomous systems across all domains, but ultimately everything in the current warfighting domains has human control at some point or another."

"None."

"None."

"Collection and display of information; algorithms to provide recommendation."

"None that are purely autonomous."

"- lane departure warning; automatic breaking; other automotive-based safety systems."

"Google maps."

"Siri on my iPhone… I think that's it."

"None."

***What characteristics of an autonomous teammate would lead you to trust or distrust it? Why?***

"I come from a community in the Navy that is beginning to implement some tools in the FCS to help operators, but is still very reliant on visual analysis of "raw" sensor data so I think some overlay or feature to view would be desired in the community I came from."

"Understanding the logic tree for decision would lead to trust. Behaviors that contradict my understanding would break trust."

"A method of relaying a confidence level on any particular data point…and sufficient time, familiarity, through training reps and sets."

"Understanding its decision logic."

"Use of data/biases."

"Does it make the right decisions; can it easily be led astray by faulty or incomplete information; how does it revert back to the HIL approach when it decides that it isn't effective by itself; does it render uncertainty accurately."

"Actions taken without "human" interface."

"- knowing how it was trained/programmed/developed   - the capability/ability to provide some explanation as to how it arrived at a conclusion or a decision   - observation of its past performance as part of the team.  - see me for results from CNO Strategic Studies Group 35 work on this exact matter."

"Duration of use, transparency of data, alignment with human experiences and expectations, alignment with mission."

"Erratic behaviors and conflicting information or recommendations would cause me to distrust it, i.e. if it makes a recommendation that wasn't even considered by me or the team, especially without a supporting justification. To trust it, I would need the opposite."

"Trust cannot be digitized. It needs to be layered. The trust inherent in an autonomous system would need to be oriented on helping me find things and helping me to create associations among massive amounts of data in the form of novel insights."

***What characteristics of autonomous teammates do you think are important to include in a human-AI teaming testbed?***

"False alerts/detections and probably just as important as missed alerts/detections. It's a very hard thing to measure in this case, but 'Did you cause no harm when compared to existing structures?'"

"Behaviors that are both expected and not expected, but correct for the situation."

"Ability to rapidly retrain from non-optimal responses."

"Ability to filter out important data."

"Human in the loop decision making."

"Ability to decipher and analyze multiple data sources simultaneously at a rapid rate."

"- the same characteristics that one would include for a traditional human teammate. We are expecting comparable performance from machines/computers/AI that we expect from human beings, so we should be using the same characteristics and standards."

"Voice interaction, alerting, explanations, urgency."

"Ability to justify recommendations, effectively communicate."

"See Rules for a Balance of Information."

***Do you have any remaining ideas, comments, or questions about autonomous teammates in the STE? If so, please record them here.***

"Autonomous teammates within the STE must be able to monitor the evolving situation and respond proactively. -This seems like it's pretty far out there. Maybe focus on more basic tasks first?...sensor correlational and fusion, mission/task assignment, status, progress reports, etc."

"These Responses will change function by function…An Autonomous Teammate will never be allowed to replace a human in the fire control loop due to legal statutes."

"I don't think that autonomy needs to replace all human functions, and I don't think it should be constrained by chain of command issues as long as it isn't in full automatic mode."

"I am not a fan of "on the loop"……and will always differ to "in the loop". Decisions of weapons, navigation and/or matters that could lead to injury and/or death need to always have a "human" component."

"- see me about the work of CNO Strategic Studies Group 35 on this matter."

"Yeah really I think if the autonomous teammate is able to act as much like a human teammate as possible, that would be ideal. If they can justify their responses/recommendations or add qualifiers (this is an opinion based on the information that I have)."

"Again, see the rules for the Balance of Information."

*Appendix B*

*Results for Domain Feature Likert Items*

"The task environment must require the same type of data gathering, decision-making, and action-taking that characterizes any military operation that employs the Observe-Orient-Decide-Act (OODA) loop."

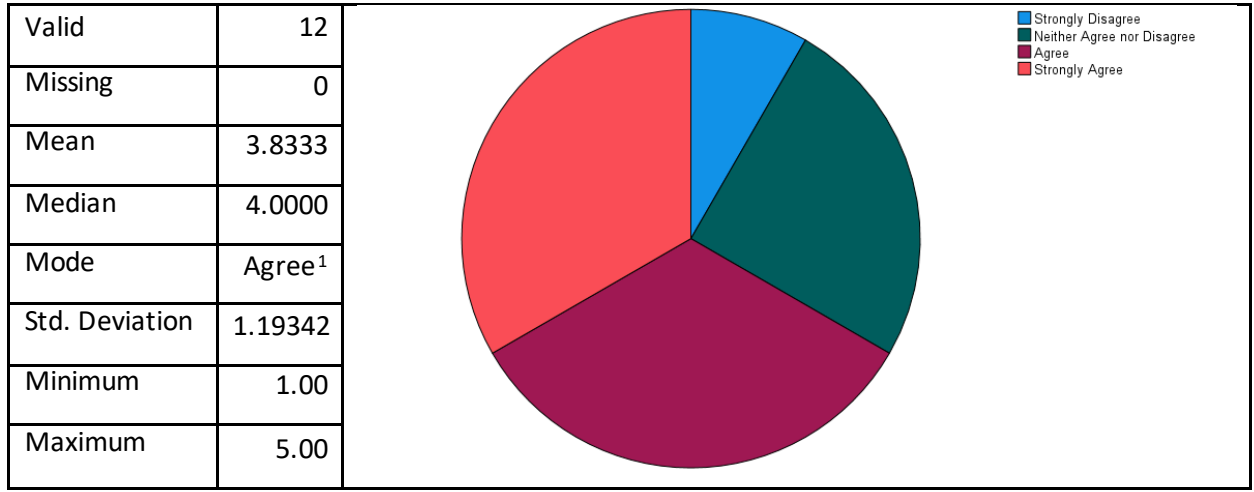

| Valid | 12 |
| --- | --- |
| Missing | 0 |
| Mean | 3.8333 |
| Median | 4.0000 |
| Mode | Agree[1] |
| Std. Deviation | 1.19342 |
| Minimum | 1.00 |
| Maximum | 5.00 |

|  |  | Frequency | Percent | Valid Percent | Cumulative Percent |
| --- | --- | --- | --- | --- | --- |
| Valid | Strongly Disagree | 1 | 8.3 | 8.3 | 8.3 |
|  | Neither Agree nor Disagree | 3 | 25.0 | 25.0 | 33.3 |
|  | Agree | 4 | 33.3 | 33.3 | 66.7 |
|  | Strongly Agree | 4 | 33.3 | 33.3 | 100.0 |
|  | Total | 12 | 100.0 | 100.0 |  |

---

[1] Multiple modes exist. The smallest value is shown.

"The task environment must require the type of coordination and decision-making anticipated within a JADC2 context."

| Valid | 12 |
|---|---|
| Missing | 0 |
| Mean | 4.5000 |
| Median | 4.5000 |
| Mode | Agree |
| Std. Deviation | .52223 |
| Minimum | 4.00 |
| Maximum | 5.00 |

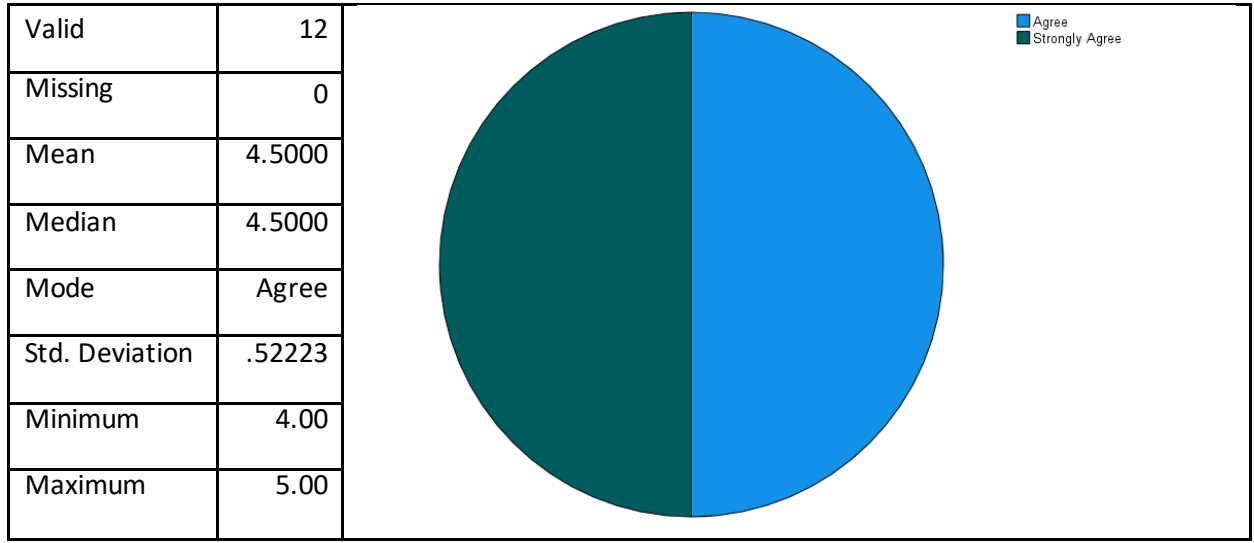

|  |  | Frequency | Percent | Valid Percent | Cumulative Percent |
|---|---|---|---|---|---|
| Valid | Agree | 6 | 50.0 | 50.0 | 50.0 |
|  | Strongly Agree | 6 | 50.0 | 50.0 | 100.0 |
|  | Total | 12 | 100.0 | 100.0 |  |

"The task environment must include support for 'transition' phases of team coordination (*e.g.*, planning sessions or AARs)."

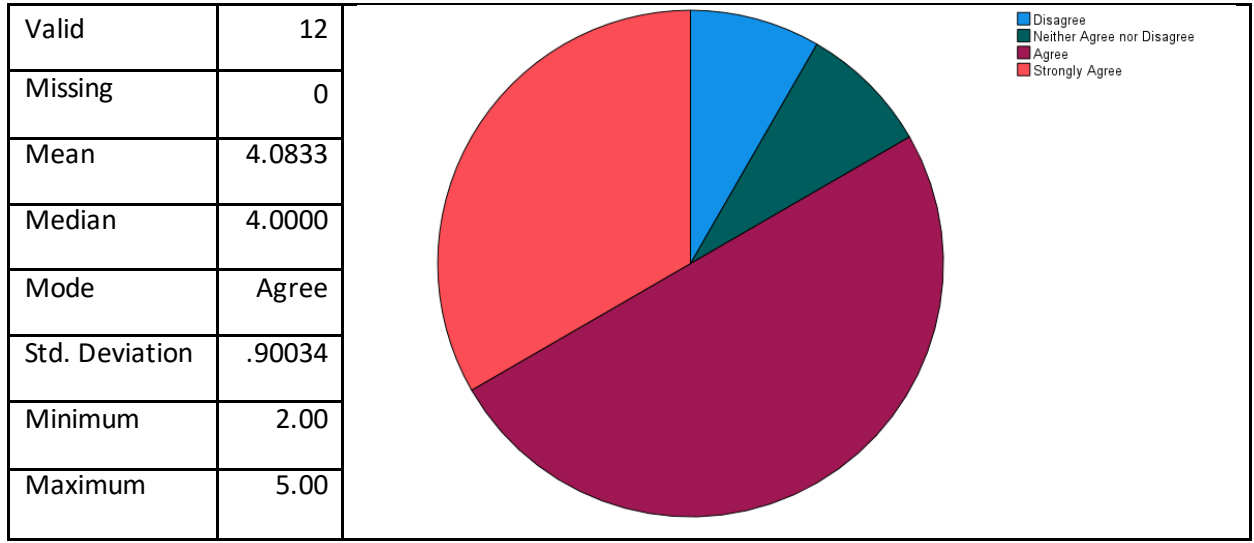

| Valid | 12 |
|---|---|
| Missing | 0 |
| Mean | 4.0833 |
| Median | 4.0000 |
| Mode | Agree |
| Std. Deviation | .90034 |
| Minimum | 2.00 |
| Maximum | 5.00 |

|  |  | Frequency | Percent | Valid Percent | Cumulative Percent |
|---|---|---|---|---|---|
| Valid | Disagree | 1 | 8.3 | 8.3 | 8.3 |
|  | Neither Agree nor Disagree | 1 | 8.3 | 8.3 | 16.7 |
|  | Agree | 6 | 50.0 | 50.0 | 66.7 |
|  | Strongly Agree | 4 | 33.3 | 33.3 | 100.0 |
|  | Total | 12 | 100.0 | 100.0 |  |

"The task environment must provide a venue for practicing and assessing coordination behaviors across the timeline (*i.e.*, from 'preparation' activities to 'adjustment' activities)."

| Valid | 12 |
|---|---|
| Missing | 0 |
| Mean | 4.3333 |
| Median | 4.5000 |
| Mode | Strongly Agree |
| Std. Deviation | .88763 |
| Minimum | 2.00 |
| Maximum | 5.00 |

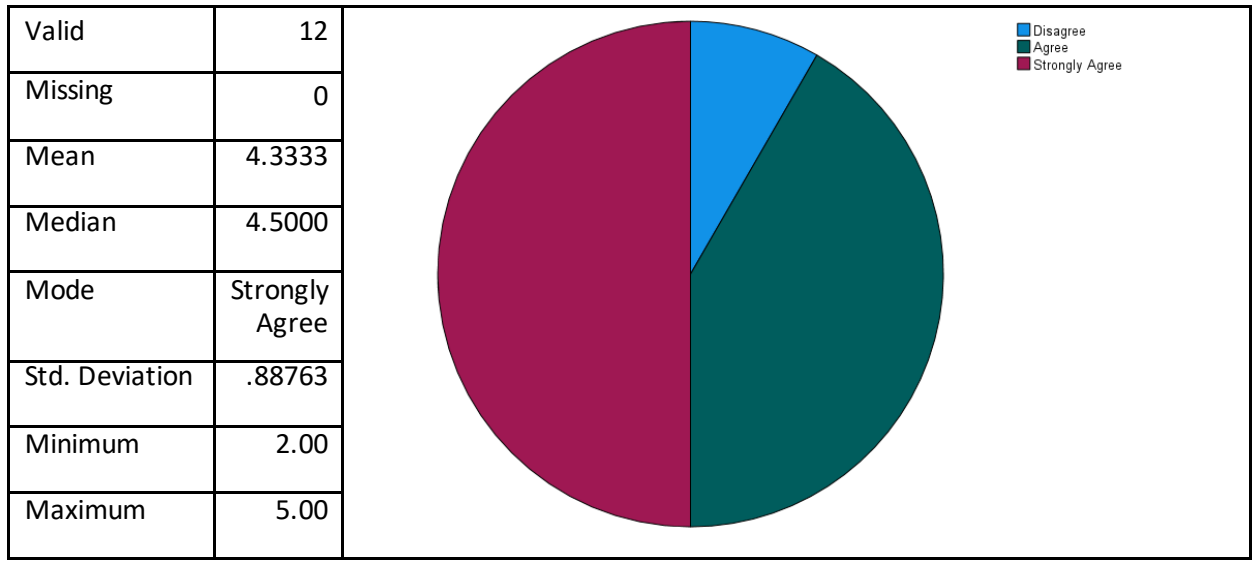

|  |  | Frequency | Percent | Valid Percent | Cumulative Percent |
|---|---|---|---|---|---|
| Valid | Disagree | 1 | 8.3 | 8.3 | 8.3 |
|  | Agree | 5 | 41.7 | 41.7 | 50.0 |
|  | Strongly Agree | 6 | 50.0 | 50.0 | 100.0 |
|  | Total | 12 | 100.0 | 100.0 |  |

"The task environment must include opportunities for mission planning, mission execution, and mission debrief."

| Valid | 12 |
|---|---|
| Missing | 0 |
| Mean | 4.3333 |
| Median | 5.0000 |
| Mode | Strongly Agree |
| Std. Deviation | .98473 |
| Minimum | 2.00 |
| Maximum | 5.00 |

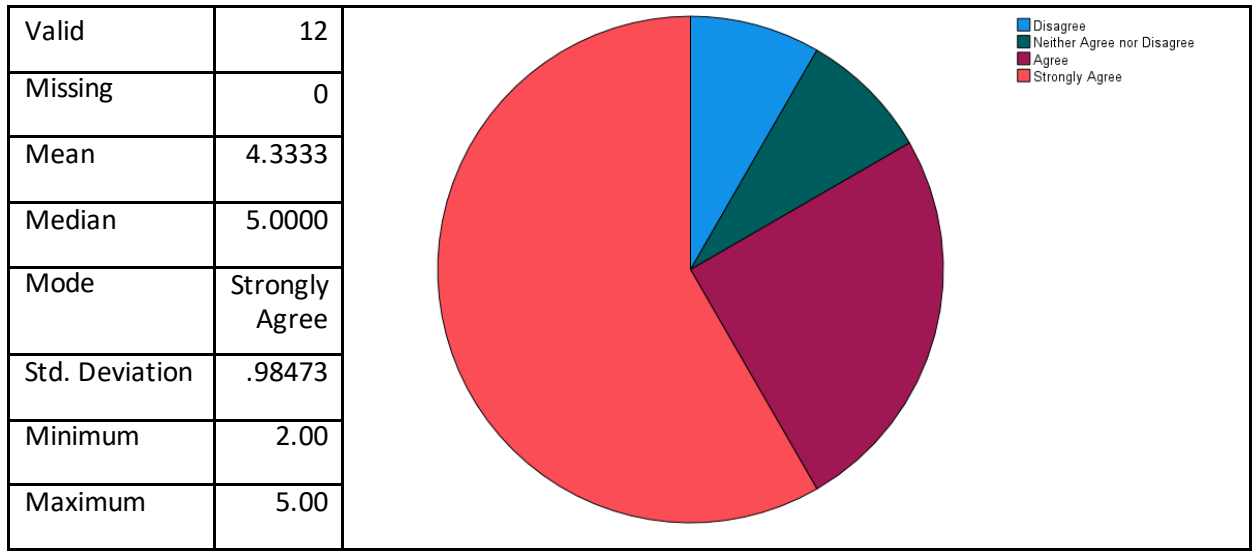

|  |  | Frequency | Percent | Valid Percent | Cumulative Percent |
|---|---|---|---|---|---|
| Valid | Disagree | 1 | 8.3 | 8.3 | 8.3 |
|  | Neither Agree nor Disagree | 1 | 8.3 | 8.3 | 16.7 |
|  | Agree | 3 | 25.0 | 25.0 | 41.7 |
|  | Strongly Agree | 7 | 58.3 | 58.3 | 100.0 |
|  | Total | 12 | 100.0 | 100.0 |  |

"The task environment must support the ability to modulate task uncertainty."

| Valid | 12 |
|---|---|
| Missing | 0 |
| Mean | 3.5833 |
| Median | 4.0000 |
| Mode | Strongly Agree |
| Std. Deviation | 1.37895 |
| Minimum | 1.00 |
| Maximum | 5.00 |

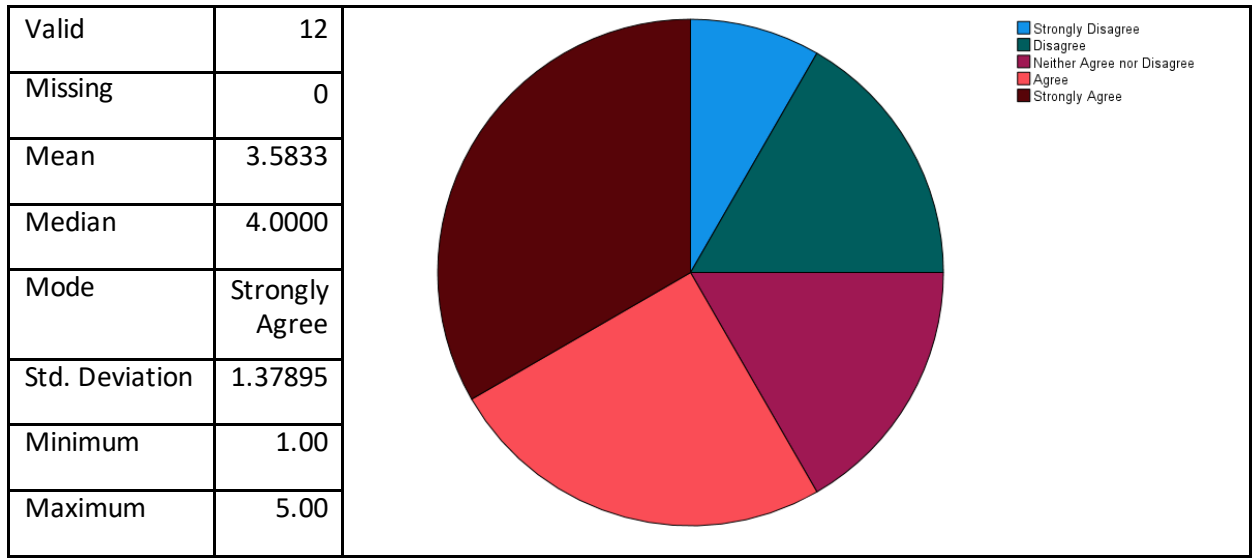

|  |  | Frequency | Percent | Valid Percent | Cumulative Percent |
|---|---|---|---|---|---|
| Valid | Strongly Disagree | 1 | 8.3 | 8.3 | 8.3 |
|  | Disagree | 2 | 16.7 | 16.7 | 25.0 |
|  | Neither Agree nor Disagree | 2 | 16.7 | 16.7 | 41.7 |
|  | Agree | 3 | 25.0 | 25.0 | 66.7 |
|  | Strongly Agree | 4 | 33.3 | 33.3 | 100.0 |
|  | Total | 12 | 100.0 | 100.0 |  |

"The tasks in the STE should have different levels of difficulty."

| Valid | 12 |
|---|---|
| Missing | 0 |
| Mean | 4.2500 |
| Median | 4.0000 |
| Mode | Agree |
| Std. Deviation | .75378 |
| Minimum | 3.00 |
| Maximum | 5.00 |

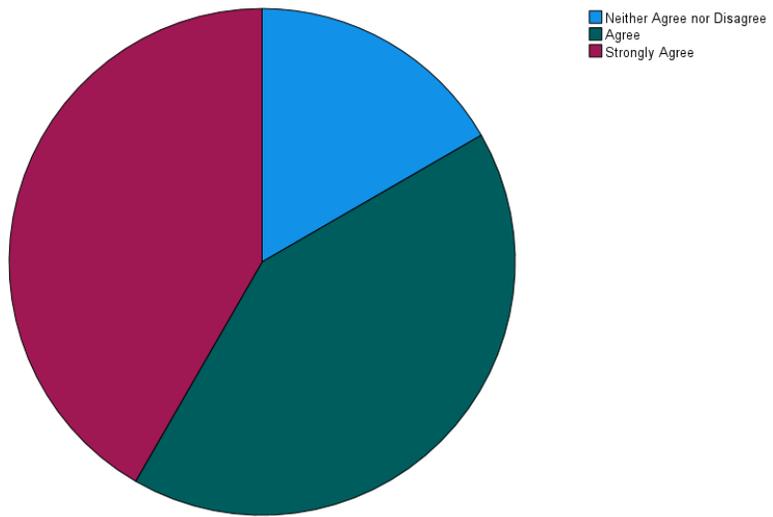

|  |  | Frequency | Percent | Valid Percent | Cumulative Percent |
|---|---|---|---|---|---|
| Valid | Neither Agree nor Disagree | 2 | 16.7 | 16.7 | 16.7 |
|  | Agree | 5 | 41.7 | 41.7 | 58.3 |
|  | Strongly Agree | 5 | 41.7 | 41.7 | 100.0 |
|  | Total | 12 | 100.0 | 100.0 |  |

"The tasks in the STE should have different levels of time pressure."

| Valid | 12 |
|---|---|
| Missing | 0 |
| Mean | 4.3333 |
| Median | 4.5000 |
| Mode | Strongly Agree |
| Std. Deviation | .77850 |
| Minimum | 3.00 |
| Maximum | 5.00 |

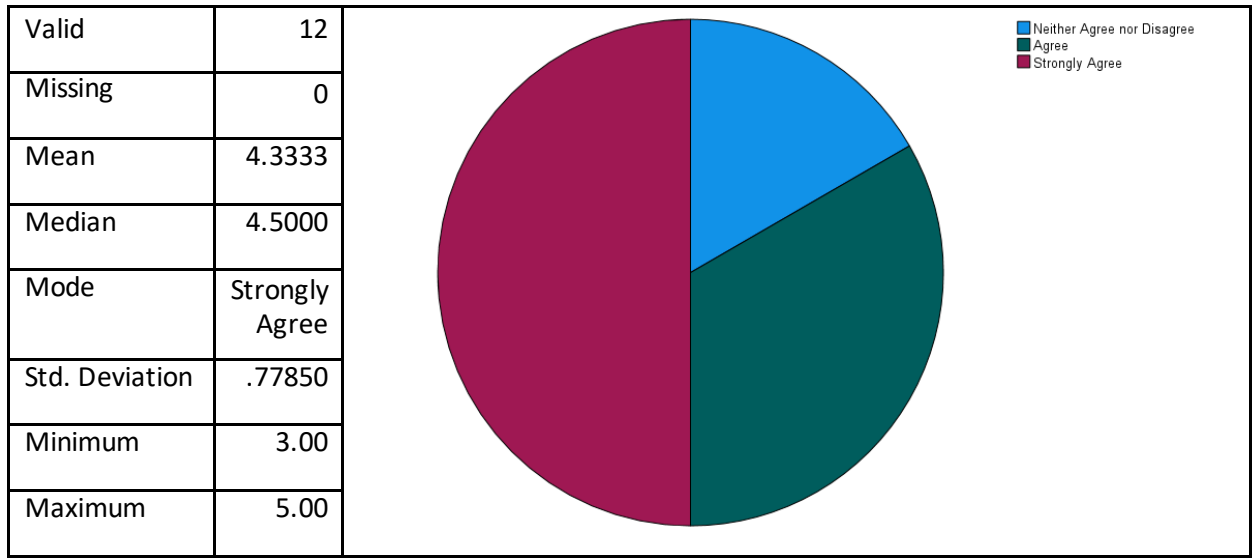

|  |  | Frequency | Percent | Valid Percent | Cumulative Percent |
|---|---|---|---|---|---|
| Valid | Neither Agree nor Disagree | 2 | 16.7 | 16.7 | 16.7 |
|  | Agree | 4 | 33.3 | 33.3 | 50.0 |
|  | Strongly Agree | 6 | 50.0 | 50.0 | 100.0 |
|  | Total | 12 | 100.0 | 100.0 |  |

"The tasks in the STE should have different levels of uncertainty."

| Valid | 12 |
|---|---|
| Missing | 0 |
| Mean | 4.0833 |
| Median | 4.5000 |
| Mode | Strongly Agree |
| Std. Deviation | 1.08362 |
| Minimum | 2.00 |
| Maximum | 5.00 |

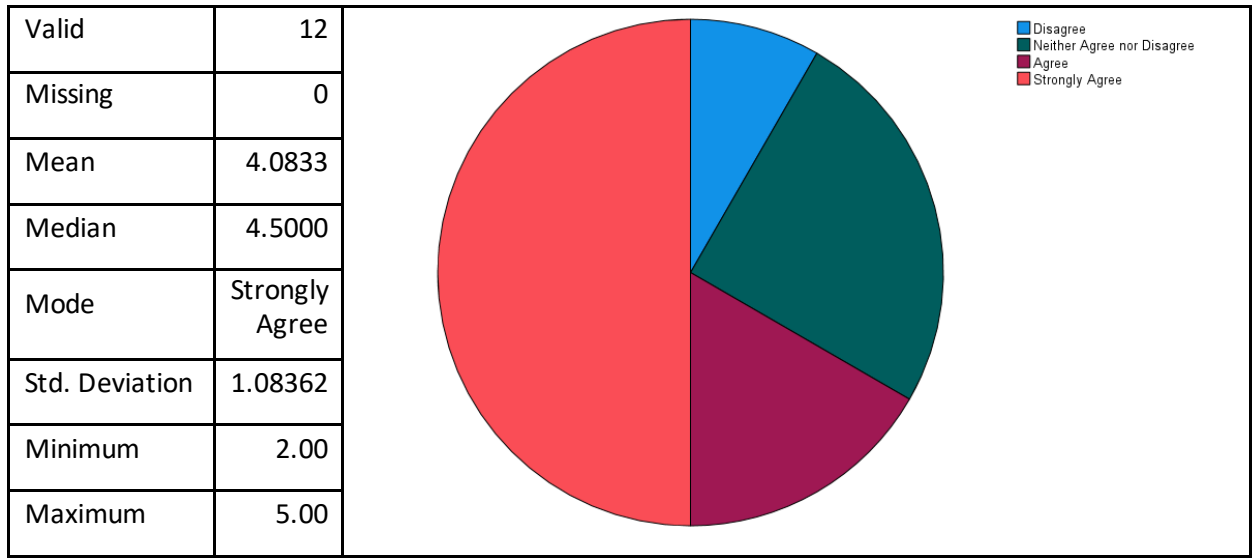

|  |  | Frequency | Percent | Valid Percent | Cumulative Percent |
|---|---|---|---|---|---|
| Valid | Disagree | 1 | 8.3 | 8.3 | 8.3 |
|  | Neither Agree nor Disagree | 3 | 25.0 | 25.0 | 33.3 |
|  | Agree | 2 | 16.7 | 16.7 | 50.0 |
|  | Strongly Agree | 6 | 50.0 | 50.0 | 100.0 |
|  | Total | 12 | 100.0 | 100.0 |  |

"The tasks in the STE should have different 'stakes' (and apply different levels of pressure on operators)."

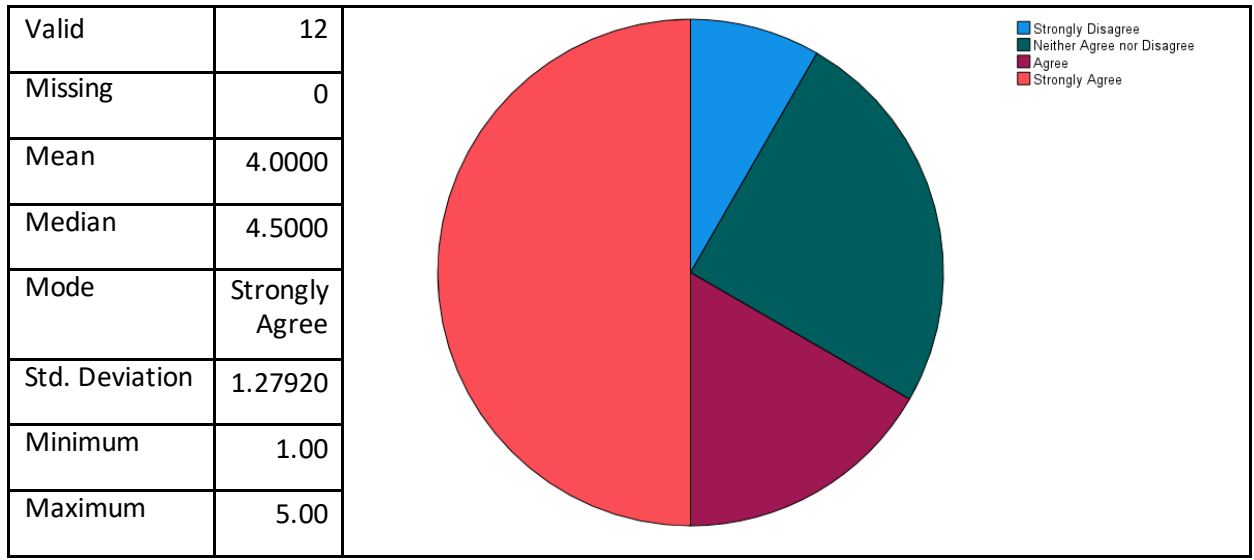

| | |
|---|---|
| Valid | 12 |
| Missing | 0 |
| Mean | 4.0000 |
| Median | 4.5000 |
| Mode | Strongly Agree |
| Std. Deviation | 1.27920 |
| Minimum | 1.00 |
| Maximum | 5.00 |

| | | Frequency | Percent | Valid Percent | Cumulative Percent |
|---|---|---|---|---|---|
| Valid | Strongly Disagree | 1 | 8.3 | 8.3 | 8.3 |
| | Neither Agree nor Disagree | 3 | 25.0 | 25.0 | 33.3 |
| | Agree | 2 | 16.7 | 16.7 | 50.0 |
| | Strongly Agree | 6 | 50.0 | 50.0 | 100.0 |
| | Total | 12 | 100.0 | 100.0 | |

This Page Intentionally Blank

*Appendix C*

**Results for Data Collection and Evaluation Likert Items**

"The STE should have a feature to provide operators with real-time feedback on their performance."

| Valid | 12 |
|---|---|
| Missing | 0 |
| Mean | 3.5833 |
| Median | 4.0000 |
| Mode | Agree |
| Std. Deviation | 1.24011 |
| Minimum | 2.00 |
| Maximum | 5.00 |

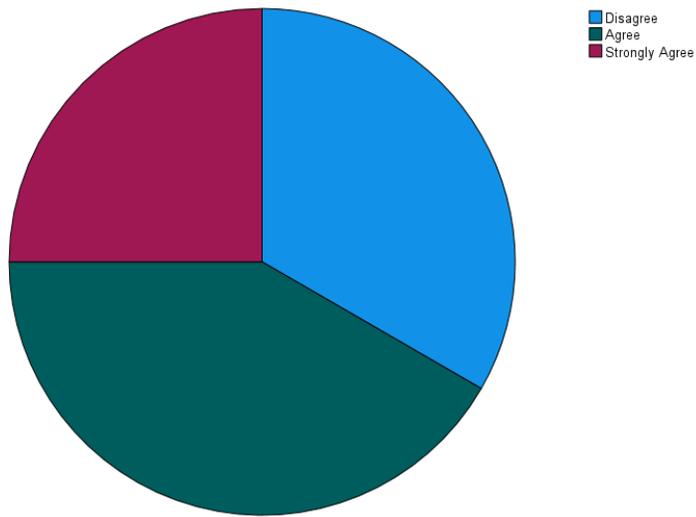

|  |  | Frequency | Percent | Valid Percent | Cumulative Percent |
|---|---|---|---|---|---|
| Valid | Disagree | 4 | 33.3 | 33.3 | 33.3 |
|  | Agree | 5 | 41.7 | 41.7 | 75.0 |
|  | Strongly Agree | 3 | 25.0 | 25.0 | 100.0 |
|  | Total | 12 | 100.0 | 100.0 |  |

"Performance evaluation in the STE should include task outcome variables (*i.e.*, if the desired outcome is achieved)."

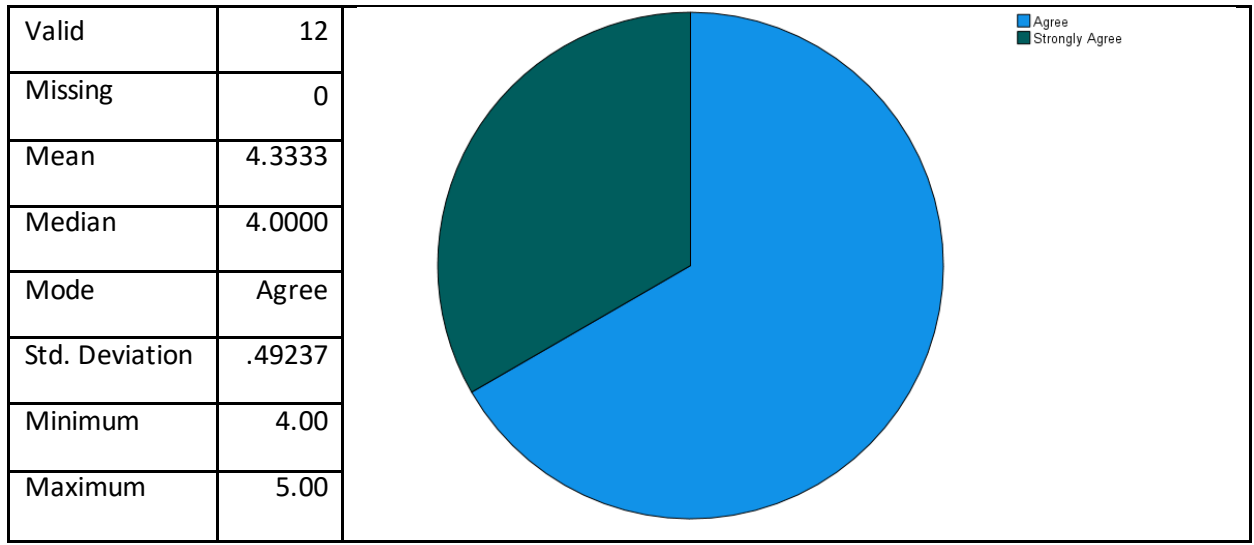

| | |
|---|---|
| Valid | 12 |
| Missing | 0 |
| Mean | 4.3333 |
| Median | 4.0000 |
| Mode | Agree |
| Std. Deviation | .49237 |
| Minimum | 4.00 |
| Maximum | 5.00 |

| | | Frequency | Percent | Valid Percent | Cumulative Percent |
|---|---|---|---|---|---|
| Valid | Agree | 8 | 66.7 | 66.7 | 66.7 |
| | Strongly Agree | 4 | 33.3 | 33.3 | 100.0 |
| | Total | 12 | 100.0 | 100.0 | |

"Performance evaluation in the STE should include task process variables (*i.e.*, WHAT actions are taken and HOW the actions are performed)."

| | |
|---|---|
| Valid | 12 |
| Missing | 0 |
| Mean | 4.6667 |
| Median | 5.0000 |
| Mode | Strongly Agree |
| Std. Deviation | .49237 |
| Minimum | 4.00 |
| Maximum | 5.00 |

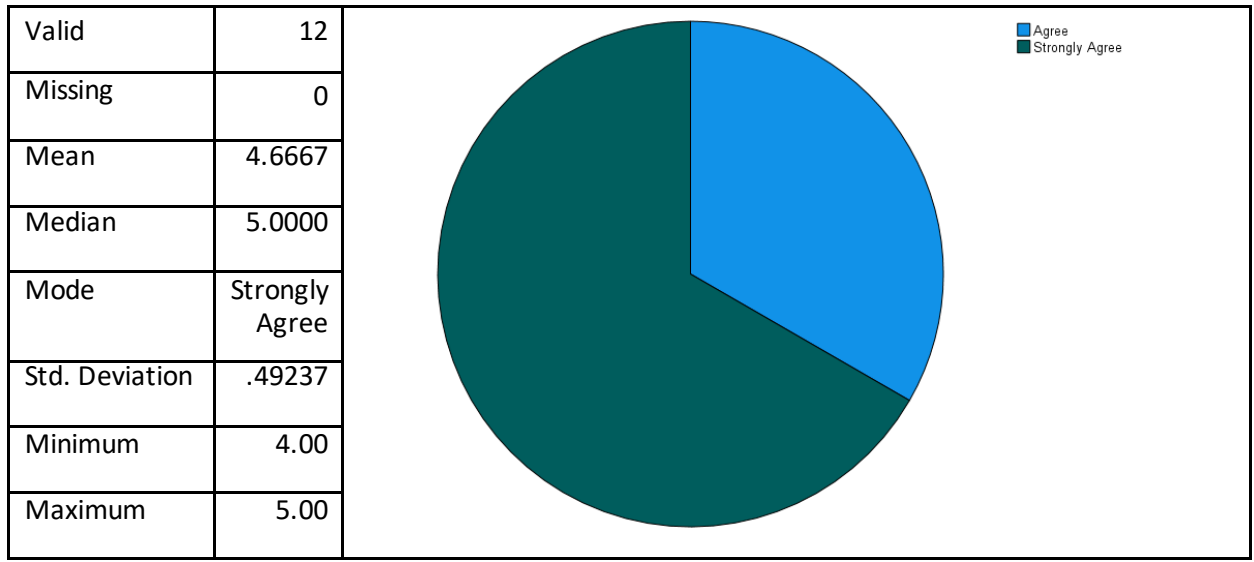

| | | Frequency | Percent | Valid Percent | Cumulative Percent |
|---|---|---|---|---|---|
| Valid | Agree | 4 | 33.3 | 33.3 | 33.3 |
| | Strongly Agree | 8 | 66.7 | 66.7 | 100.0 |
| | Total | 12 | 100.0 | 100.0 | |

"Performance evaluation in the STE should include how well the team members coordinate with one another."

| Valid | 12 |
|---|---|
| Missing | 0 |
| Mean | 4.3333 |
| Median | 5.0000 |
| Mode | Strongly Agree |
| Std. Deviation | .88763 |
| Minimum | 3.00 |
| Maximum | 5.00 |

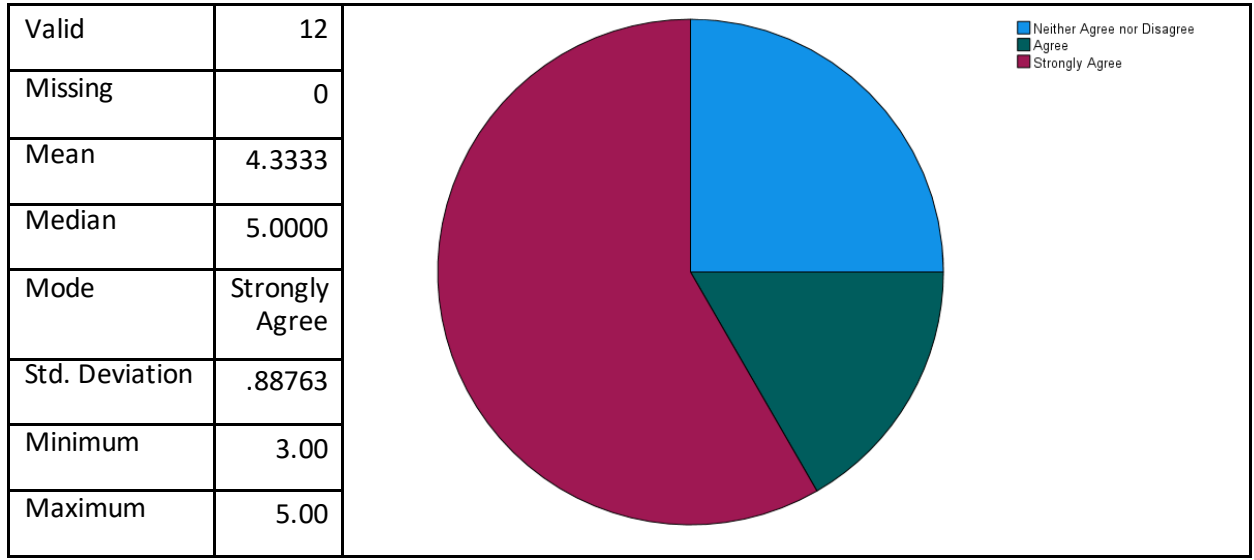

|  |  | Frequency | Percent | Valid Percent | Cumulative Percent |
|---|---|---|---|---|---|
| Valid | Neither Agree nor Disagree | 3 | 25.0 | 25.0 | 25.0 |
|  | Agree | 2 | 16.7 | 16.7 | 41.7 |
|  | Strongly Agree | 7 | 58.3 | 58.3 | 100.0 |
|  | Total | 12 | 100.0 | 100.0 |  |

This Page Intentionally Blank

*Appendix D*

*Results for Communication-focused Likert Items*

"The STE must enable spoken communication among teammates."

| Valid | 12 |
|---|---|
| Missing | 0 |
| Mean | 4.3333 |
| Median | 5.0000 |
| Mode | Strongly Agree |
| Std. Deviation | 1.15470 |
| Minimum | 1.00 |
| Maximum | 5.00 |

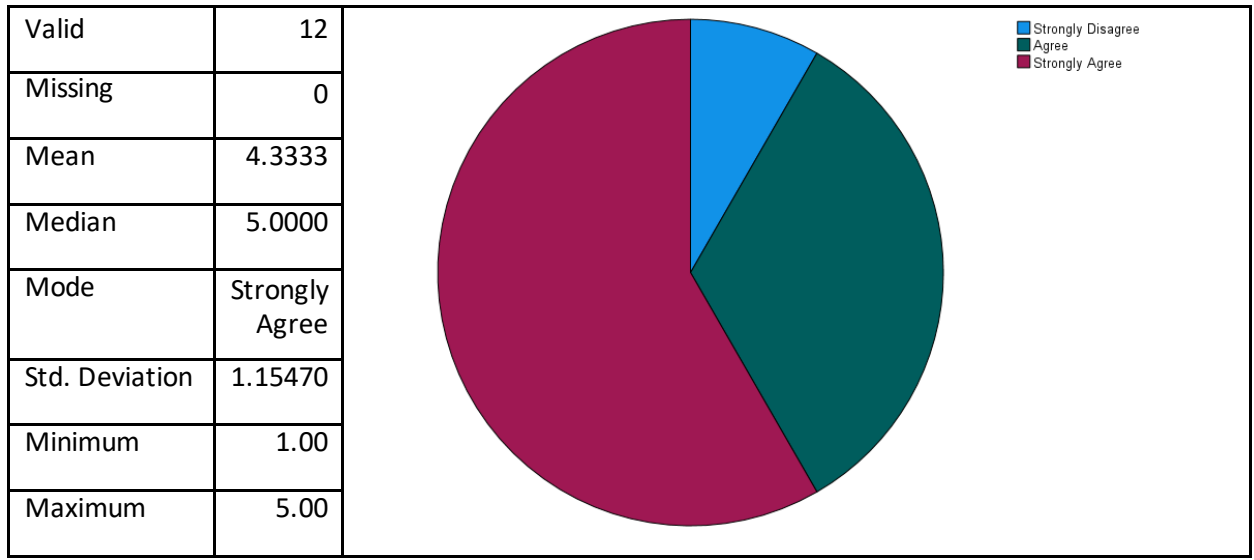

|  |  | Frequency | Percent | Valid Percent | Cumulative Percent |
|---|---|---|---|---|---|
| Valid | Strongly Disagree | 1 | 8.3 | 8.3 | 8.3 |
|  | Agree | 4 | 33.3 | 33.3 | 41.7 |
|  | Strongly Agree | 7 | 58.3 | 58.3 | 100.0 |
|  | Total | 12 | 100.0 | 100.0 |  |

"The STE must enable chat-based communication among teammates."

| Valid | 12 |
|---|---|
| Missing | 0 |
| Mean | 4.2500 |
| Median | 4.0000 |
| Mode | Agree |
| Std. Deviation | .75378 |
| Minimum | 3.00 |
| Maximum | 5.00 |

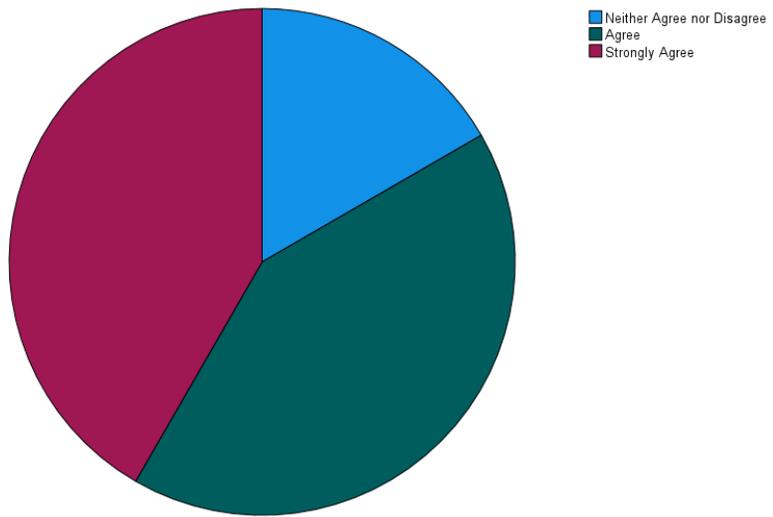

|  |  | Frequency | Percent | Valid Percent | Cumulative Percent |
|---|---|---|---|---|---|
| Valid | Neither Agree nor Disagree | 2 | 16.7 | 16.7 | 16.7 |
|  | Agree | 5 | 41.7 | 41.7 | 58.3 |
|  | Strongly Agree | 5 | 41.7 | 41.7 | 100.0 |
|  | Total | 12 | 100.0 | 100.0 |  |

"The STE must enable email communication among teammates."

| | |
|---|---|
| Valid | 12 |
| Missing | 0 |
| Mean | 4.0000 |
| Median | 4.0000 |
| Mode | Agree |
| Std. Deviation | .73855 |
| Minimum | 3.00 |
| Maximum | 5.00 |

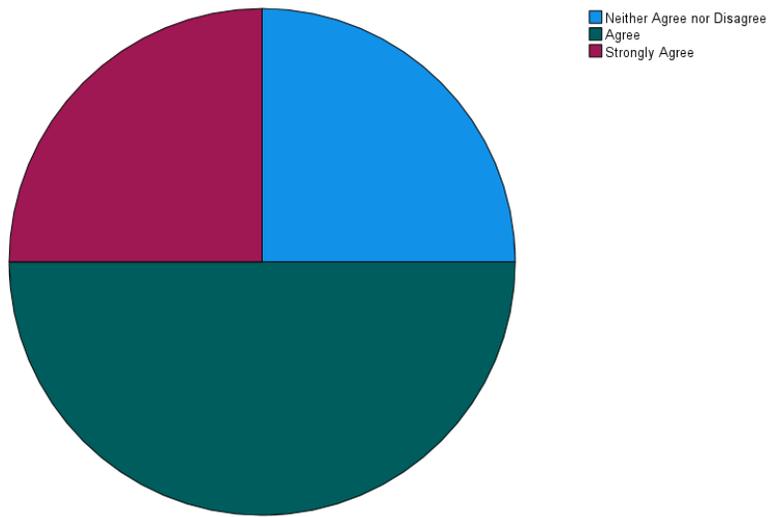

| | | Frequency | Percent | Valid Percent | Cumulative Percent |
|---|---|---|---|---|---|
| Valid | Neither Agree nor Disagree | 3 | 25.0 | 25.0 | 25.0 |
| | Agree | 6 | 50.0 | 50.0 | 75.0 |
| | Strongly Agree | 3 | 25.0 | 25.0 | 100.0 |
| | Total | 12 | 100.0 | 100.0 | |

"The STE must enable video communication among teammates."

| | |
|---|---|
| Valid | 12 |
| Missing | 0 |
| Mean | 3.5833 |
| Median | 4.0000 |
| Mode | Agree |
| Std. Deviation | 1.08362 |
| Minimum | 2.00 |
| Maximum | 5.00 |

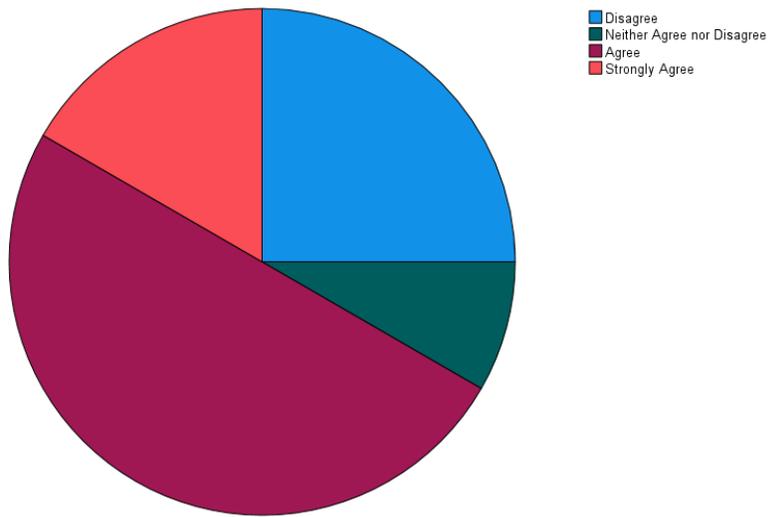

| | | Frequency | Percent | Valid Percent | Cumulative Percent |
|---|---|---|---|---|---|
| Valid | Disagree | 3 | 25.0 | 25.0 | 25.0 |
| | Neither Agree nor Disagree | 1 | 8.3 | 8.3 | 33.3 |
| | Agree | 6 | 50.0 | 50.0 | 83.3 |
| | Strongly Agree | 2 | 16.7 | 16.7 | 100.0 |
| | Total | 12 | 100.0 | 100.0 | |

"The STE must enable paper-based communication among teammates."

| Valid | 12 |
|---|---|
| Missing | 0 |
| Mean | 3.0000 |
| Median | 3.0000 |
| Mode | Agree |
| Std. Deviation | 1.04447 |
| Minimum | 1.00 |
| Maximum | 4.00 |

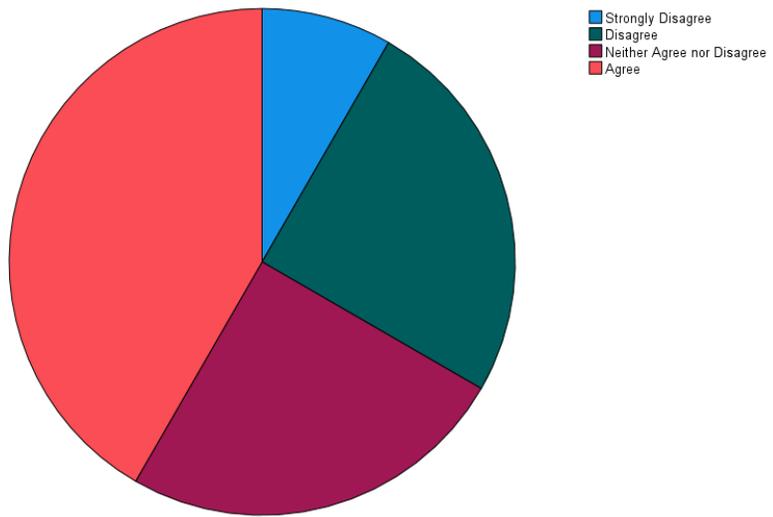

|  |  | Frequency | Percent | Valid Percent | Cumulative Percent |
|---|---|---|---|---|---|
| Valid | Strongly Disagree | 1 | 8.3 | 8.3 | 8.3 |
|  | Disagree | 3 | 25.0 | 25.0 | 33.3 |
|  | Neither Agree nor Disagree | 3 | 25.0 | 25.0 | 58.3 |
|  | Agree | 5 | 41.7 | 41.7 | 100.0 |
|  | Total | 12 | 100.0 | 100.0 |  |

*Appendix E*

*Results for Agent-related Likert Items*

"Within the STE, the roles that humans fill and the roles that autonomous teammates fill should be interchangeable."

| Valid | 12 |
|---|---|
| Missing | 0 |
| Mean | 2.5000 |
| Median | 2.0000 |
| Mode | Disagree |
| Std. Deviation | 1.08711 |
| Minimum | 1.00 |
| Maximum | 5.00 |

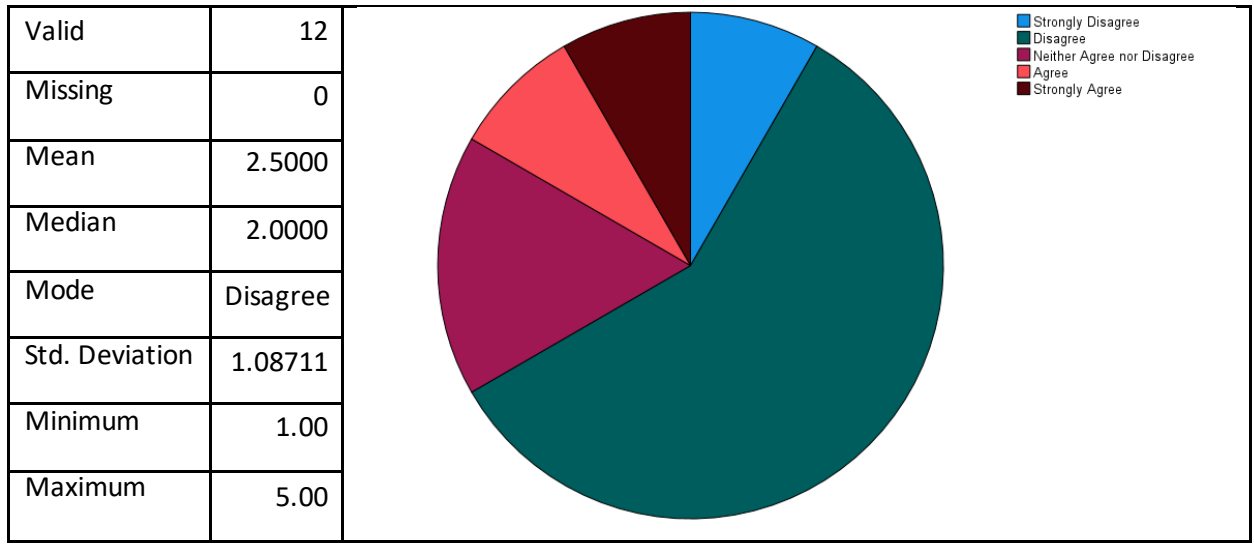

| | | Frequency | Percent | Valid Percent | Cumulative Percent |
|---|---|---|---|---|---|
| Valid | Strongly Disagree | 1 | 8.3 | 8.3 | 8.3 |
| | Disagree | 7 | 58.3 | 58.3 | 66.7 |
| | Neither Agree nor Disagree | 2 | 16.7 | 16.7 | 83.3 |
| | Agree | 1 | 8.3 | 8.3 | 91.7 |
| | Strongly Agree | 1 | 8.3 | 8.3 | 100.0 |
| | Total | 12 | 100.0 | 100.0 | |

"Autonomous teammates within the STE must be able to monitor the state of the mission, task, and sub-tasks the team is performing."

| Valid | 12 |
|---|---|
| Missing | 0 |
| Mean | 3.9167 |
| Median | 4.0000 |
| Mode | Agree |
| Std. Deviation | .51493 |
| Minimum | 3.00 |
| Maximum | 5.00 |

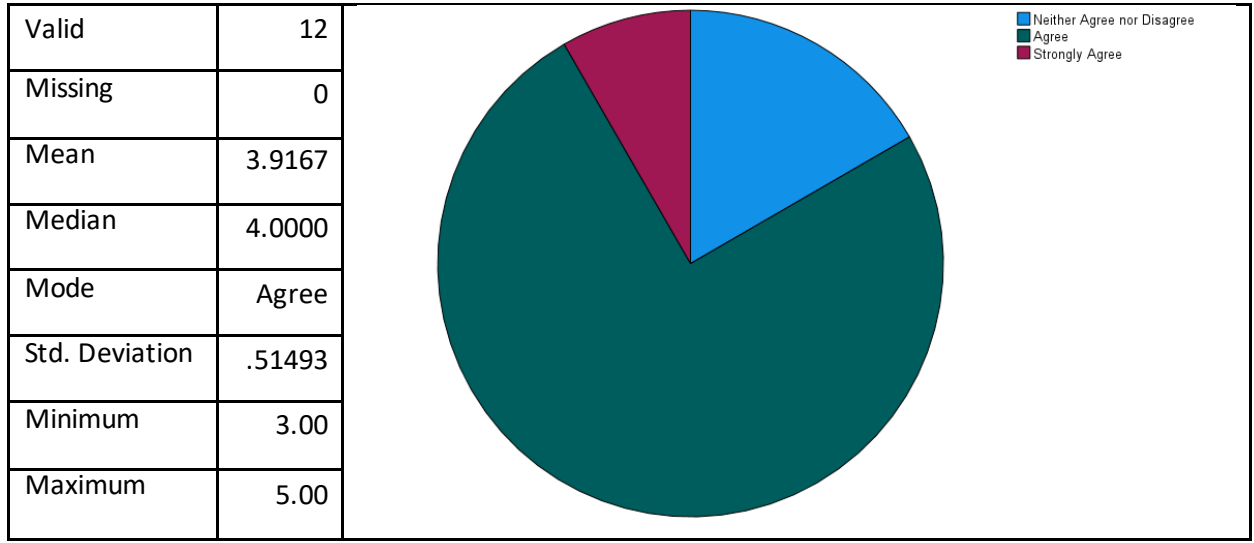

|  |  | Frequency | Percent | Valid Percent | Cumulative Percent |
|---|---|---|---|---|---|
| Valid | Neither Agree nor Disagree | 2 | 16.7 | 16.7 | 16.7 |
|  | Agree | 9 | 75.0 | 75.0 | 91.7 |
|  | Strongly Agree | 1 | 8.3 | 8.3 | 100.0 |
|  | Total | 12 | 100.0 | 100.0 |  |

"Autonomous teammates within the STE must be able to perform all activities normally associated with the role to which they are assigned."

| Valid | 12 |
|---|---|
| Missing | 0 |
| Mean | 4.0833 |
| Median | 4.0000 |
| Mode | Strongly Agree |
| Std. Deviation | .99620 |
| Minimum | 2.00 |
| Maximum | 5.00 |

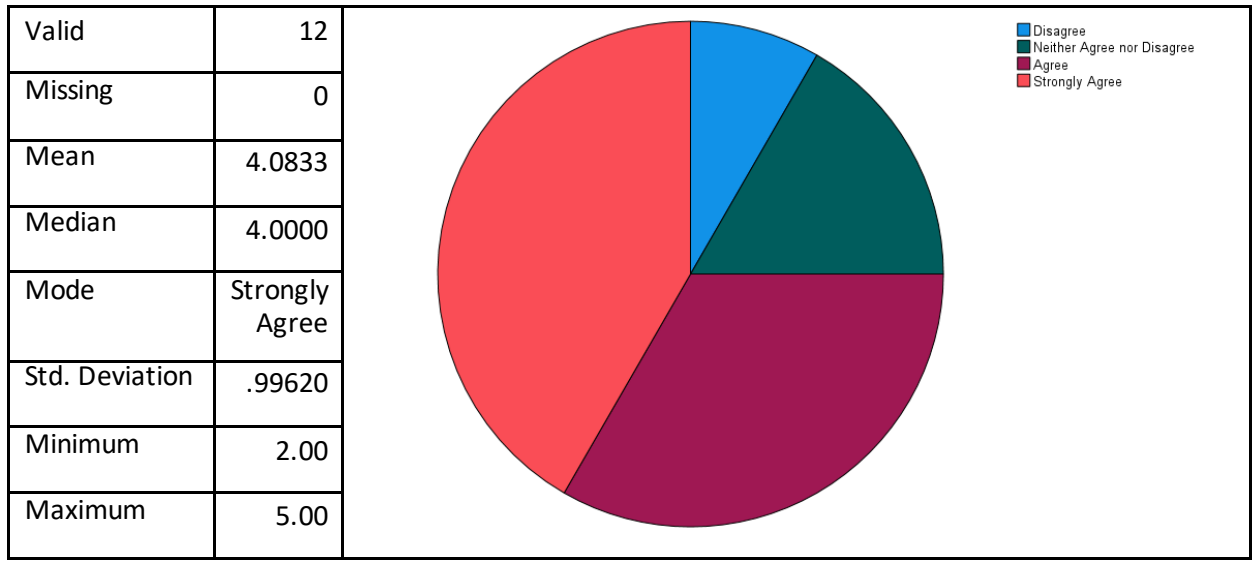

|  |  | Frequency | Percent | Valid Percent | Cumulative Percent |
|---|---|---|---|---|---|
| Valid | Disagree | 1 | 8.3 | 8.3 | 8.3 |
|  | Neither Agree nor Disagree | 2 | 16.7 | 16.7 | 25.0 |
|  | Agree | 4 | 33.3 | 33.3 | 58.3 |
|  | Strongly Agree | 5 | 41.7 | 41.7 | 100.0 |
|  | Total | 12 | 100.0 | 100.0 |  |

"Autonomous teammates within the STE must be able to accept guidance from teammates (human or non-human) as conditions and priorities change."

| Valid | 12 |
|---|---|
| Missing | 0 |
| Mean | 4.5000 |
| Median | 5.0000 |
| Mode | Strongly Agree |
| Std. Deviation | .90453 |
| Minimum | 2.00 |
| Maximum | 5.00 |

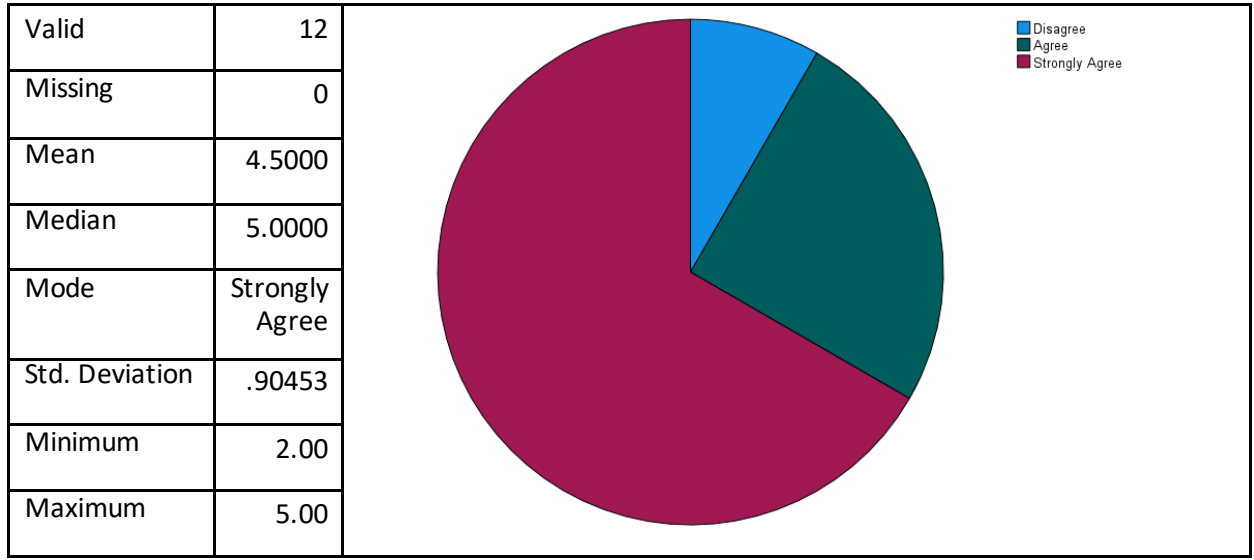

|  |  | Frequency | Percent | Valid Percent | Cumulative Percent |
|---|---|---|---|---|---|
| Valid | Disagree | 1 | 8.3 | 8.3 | 8.3 |
|  | Agree | 3 | 25.0 | 25.0 | 33.3 |
|  | Strongly Agree | 8 | 66.7 | 66.7 | 100.0 |
|  | Total | 12 | 100.0 | 100.0 |  |

"Autonomous teammates within the STE must be able to monitor the evolving situation and respond proactively."

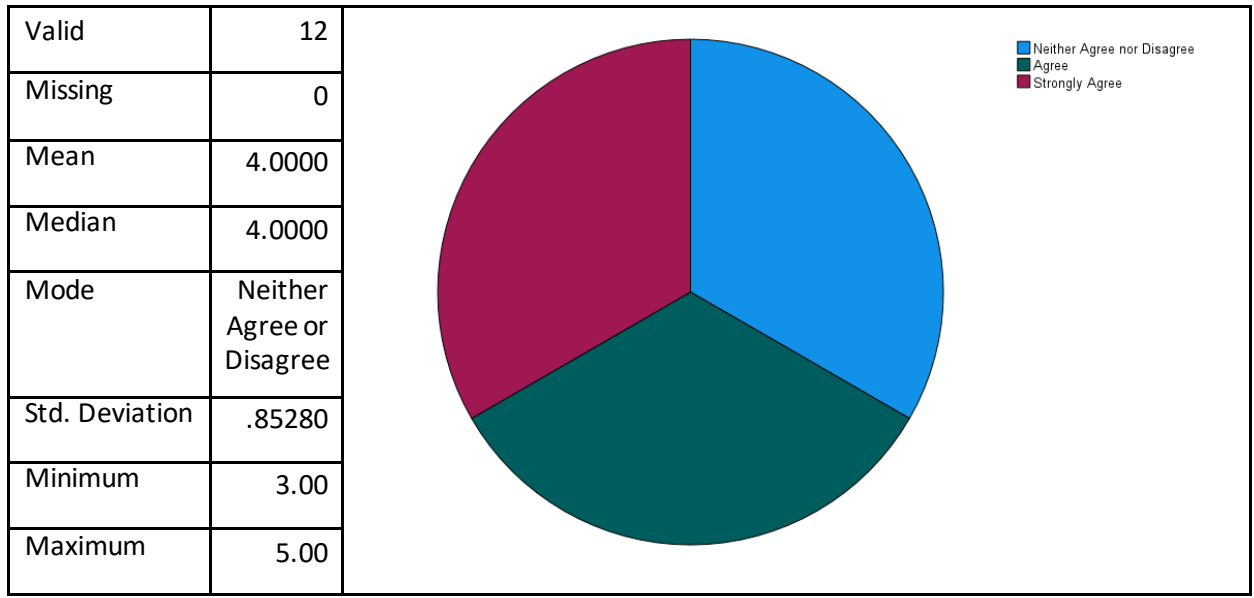

| Valid | 12 |
|---|---|
| Missing | 0 |
| Mean | 4.0000 |
| Median | 4.0000 |
| Mode | Neither Agree or Disagree |
| Std. Deviation | .85280 |
| Minimum | 3.00 |
| Maximum | 5.00 |

|  |  | Frequency | Percent | Valid Percent | Cumulative Percent |
|---|---|---|---|---|---|
| Valid | Neither Agree nor Disagree | 4 | 33.3 | 33.3 | 33.3 |
|  | Agree | 4 | 33.3 | 33.3 | 66.7 |
|  | Strongly Agree | 4 | 33.3 | 33.3 | 100.0 |
|  | Total | 12 | 100.0 | 100.0 |  |

"Autonomous teammates within the STE must be able to adhere to a chain of command."

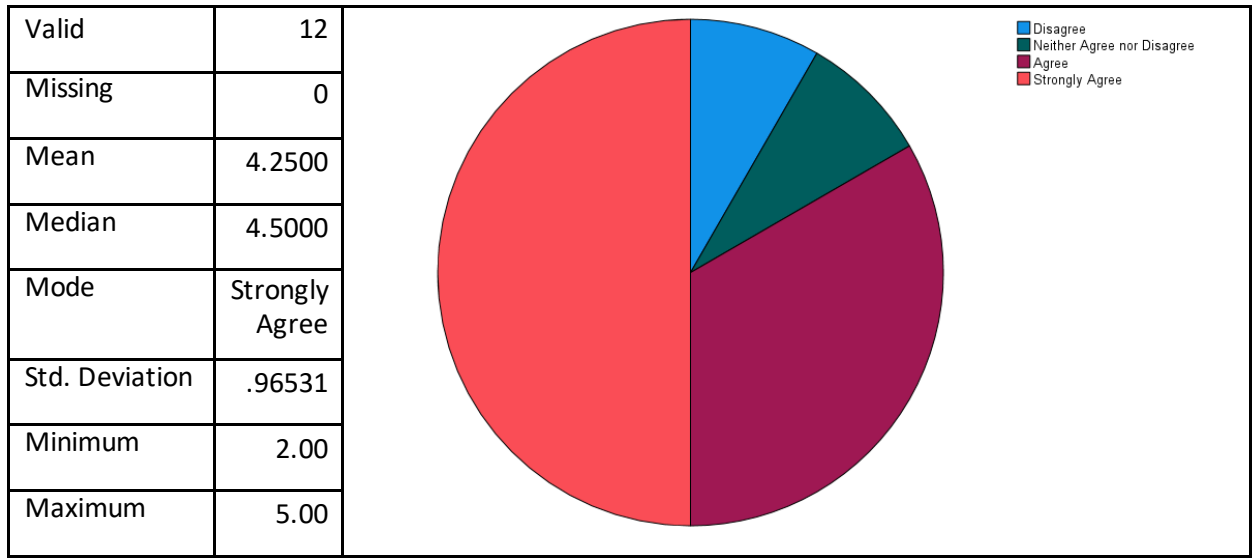

| Valid | 12 |
|---|---|
| Missing | 0 |
| Mean | 4.2500 |
| Median | 4.5000 |
| Mode | Strongly Agree |
| Std. Deviation | .96531 |
| Minimum | 2.00 |
| Maximum | 5.00 |

|  |  | Frequency | Percent | Valid Percent | Cumulative Percent |
|---|---|---|---|---|---|
| Valid | Disagree | 1 | 8.3 | 8.3 | 8.3 |
|  | Neither Agree nor Disagree | 1 | 8.3 | 8.3 | 16.7 |
|  | Agree | 4 | 33.3 | 33.3 | 50.0 |
|  | Strongly Agree | 6 | 50.0 | 50.0 | 100.0 |
|  | Total | 12 | 100.0 | 100.0 |  |

"Autonomous teammates within the STE must be able to notify teammates of on-going progress, state changes, etc."

| Valid | 12 |
|---|---|
| Missing | 0 |
| Mean | 4.6667 |
| Median | 5.0000 |
| Mode | Strongly Agree |
| Std. Deviation | .49237 |
| Minimum | 4.00 |
| Maximum | 5.00 |

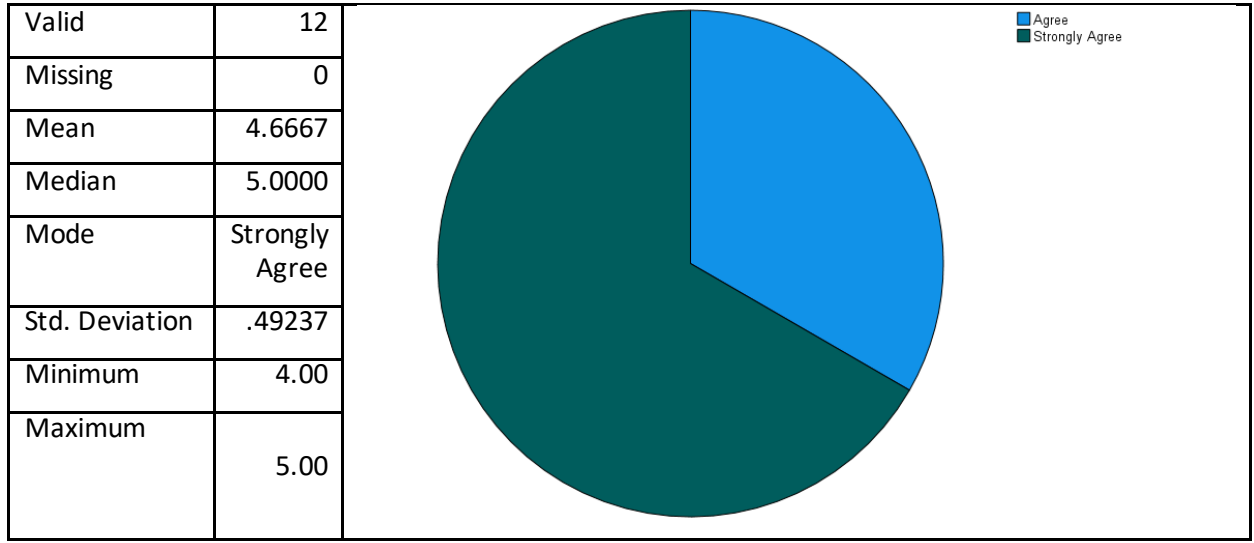

| | | Frequency | Percent | Valid Percent | Cumulative Percent |
|---|---|---|---|---|---|
| Valid | Agree | 4 | 33.3 | 33.3 | 33.3 |
| | Strongly Agree | 8 | 66.7 | 66.7 | 100.0 |
| | Total | 12 | 100.0 | 100.0 | |

"Autonomous teammates within the STE must be able to acknowledge inputs from other teammates."

| | |
|---|---|
| Valid | 12 |
| Missing | 0 |
| Mean | 4.2500 |
| Median | 4.0000 |
| Mode | Agree |
| Std. Deviation | .75378 |
| Minimum | 3.00 |
| Maximum | 5.00 |

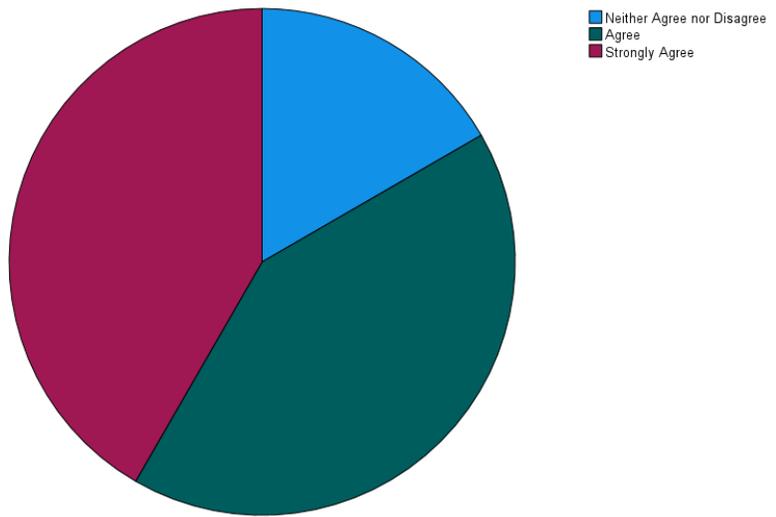

| | | Frequency | Percent | Valid Percent | Cumulative Percent |
|---|---|---|---|---|---|
| Valid | Neither Agree nor Disagree | 2 | 16.7 | 16.7 | 16.7 |
| | Agree | 5 | 41.7 | 41.7 | 58.3 |
| | Strongly Agree | 5 | 41.7 | 41.7 | 100.0 |
| | Total | 12 | 100.0 | 100.0 | |

"Autonomous teammates within the STE must be able to repair trust violations resulting from errors."

| Valid | 12 |
|---|---|
| Missing | 0 |
| Mean | 3.9167 |
| Median | 4.0000 |
| Mode | Agree |
| Std. Deviation | .79296 |
| Minimum | 3.00 |
| Maximum | 5.00 |

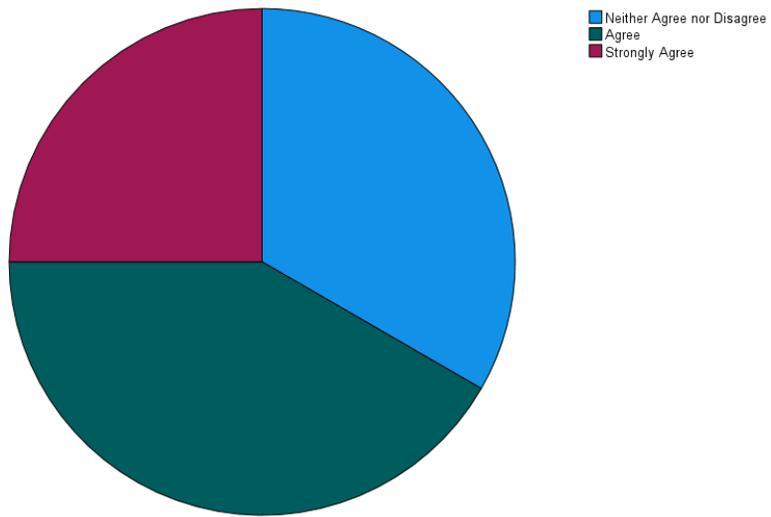

| | | Frequency | Percent | Valid Percent | Cumulative Percent |
|---|---|---|---|---|---|
| Valid | Neither Agree nor Disagree | 4 | 33.3 | 33.3 | 33.3 |
| | Agree | 5 | 41.7 | 41.7 | 75.0 |
| | Strongly Agree | 3 | 25.0 | 25.0 | 100.0 |
| | Total | 12 | 100.0 | 100.0 | |

"Autonomous teammates within the STE must be able to temper trust by actively monitoring for conditions in which failure was likely and notifying teammates that they should not rely on the automation in such situations in the future."

| | |
|---|---|
| Valid | 12 |
| Missing | 0 |
| Mean | 4.2500 |
| Median | 4.5000 |
| Mode | Strongly Agree |
| Std. Deviation | .86603 |
| Minimum | 3.00 |
| Maximum | 5.00 |

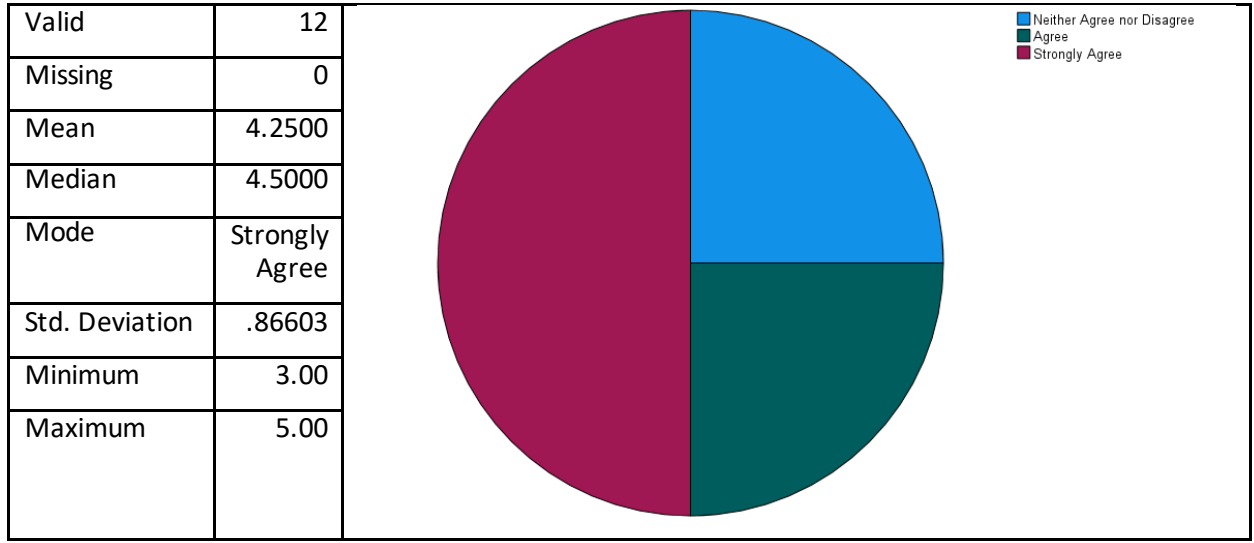

| | | Frequency | Percent | Valid Percent | Cumulative Percent |
|---|---|---|---|---|---|
| Valid | Neither Agree nor Disagree | 3 | 25.0 | 25.0 | 25.0 |
| | Agree | 3 | 25.0 | 25.0 | 50.0 |
| | Strongly Agree | 6 | 50.0 | 50.0 | 100.0 |
| | Total | 12 | 100.0 | 100.0 | |

"Autonomous teammates within the STE must be able to accurately monitor and report their status."

| Valid | 12 |
|---|---|
| Missing | 0 |
| Mean | 4.2500 |
| Median | 4.0000 |
| Mode | Agree |
| Std. Deviation | .86603 |
| Minimum | 2.00 |
| Maximum | 5.00 |

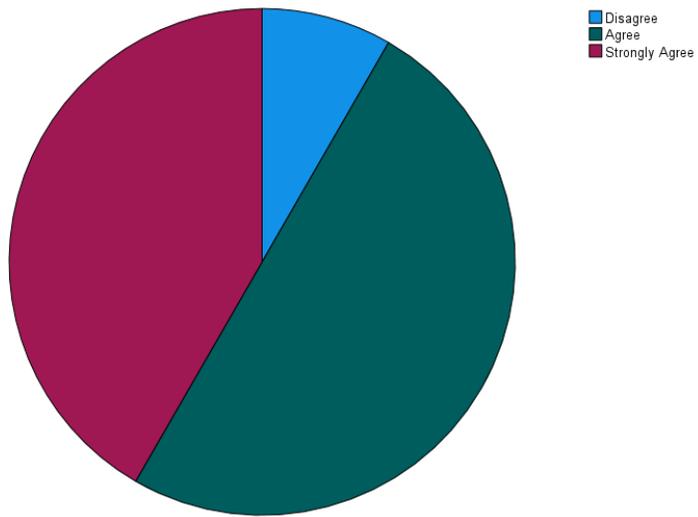

|  |  | Frequency | Percent | Valid Percent | Cumulative Percent |
|---|---|---|---|---|---|
| Valid | Disagree | 1 | 8.3 | 8.3 | 8.3 |
|  | Agree | 6 | 50.0 | 50.0 | 58.3 |
|  | Strongly Agree | 5 | 41.7 | 41.7 | 100.0 |
|  | Total | 12 | 100.0 | 100.0 |  |